\documentclass[11pt,a4paper]{article}

\usepackage{amssymb,amsmath}
\usepackage{ifxetex,ifluatex}
\usepackage{fixltx2e} 
\PassOptionsToPackage{hyphens}{url} 
\usepackage[unicode=true]{hyperref}
\hypersetup{
  pdftitle={FFORMPP: Feature-based forecast model performance prediction},
  pdfkeywords={Time series, meta-learning, mixture autoregressive models, surface regression, M4 Competition},
  pdfborder={0 0 0},
  breaklinks=true}
\usepackage{geometry}
\usepackage{rotating}

\usepackage[style=authoryear-comp, backend=biber, natbib=true]{biblatex}
\usepackage{longtable,booktabs}
\usepackage{graphicx,grffile}
\makeatletter
\def\maxwidth{\ifdim\Gin@nat@width>\linewidth\linewidth\else\Gin@nat@width\fi}
\def\maxheight{\ifdim\Gin@nat@height>\textheight\textheight\else\Gin@nat@height\fi}
\makeatother
\setkeys{Gin}{width=\maxwidth,height=\maxheight,keepaspectratio}

\makeatletter
\def\fps@figure{htbp}
\makeatother

\let\pkg=\texttt
\let\proglang=\textsf

\title{FFORMPP: Feature-based forecast model performance prediction}

\author{Thiyanga S. Talagala\footnote{Department of Econometrics and Business Statistics,
    Monash University, VIC 3800, Australia, and Department of Statistics, Faculty of
    Applied Sciences, University of Sri Jayewardenepura, Colombo, Sri Lanka.  Email:
    \url{ttalagala@sjp.ac.lk}. ORCID: 0000-0002-0656-9789.}, Feng Li\footnote{School of Statistics and Mathematics, Central
    University of Finance and Economics, Beijing 102206, China. Email:
    \url{feng.li@cufe.edu.cn}. ORCID: 0000-0002-4248-9778.}, Yanfei Kang\footnote{School of Economics and Management,
    Beihang University, Beijing 100191, China. Email:
    \url{yanfeikang@buaa.edu.cn}. ORCID: 0000-0001-8769-6650. Corresponding author.}}

\RequirePackage{caption}
\DeclareCaptionStyle{italic}[justification=centering]
{labelfont={bf},textfont={it},labelsep=colon}
\RequirePackage{bm,amsmath}
\allowdisplaybreaks

\RequirePackage{graphicx}
\RequirePackage{setspace}
\spacing{1.5}

\RequirePackage{microtype}

\usepackage{indentfirst}

\RequirePackage{xcolor} 
\definecolor{darkblue}{rgb}{0,0,.6}
\RequirePackage{url}

\makeatletter
\@ifpackageloaded{hyperref}{}{\RequirePackage{hyperref}}
\makeatother
\hypersetup{
  citecolor=0 0 0,
  breaklinks=true,
  bookmarksopen=true,
  bookmarksnumbered=true,
  linkcolor=darkblue,
  urlcolor=blue,
  citecolor=darkblue,
  colorlinks=true}

\newenvironment{keywords}{\par\vspace{0.5cm}\noindent{\textbf{Keywords:}}}{\vspace{0.25cm}\par\vspace{0.5cm}\par}

\usepackage[T1]{fontenc}
\usepackage[utf8]{inputenc}

\usepackage[showonlyrefs]{mathtools}
\usepackage[no-weekday]{eukdate}

\makeatother
\DeclareFieldFormat{url}{\texttt{\url{#1}}}
\DeclareFieldFormat[article]{pages}{#1}
\DeclareFieldFormat[inproceedings]{pages}{\lowercase{pp.}#1}
\DeclareFieldFormat[incollection]{pages}{\lowercase{pp.}#1}
\DeclareFieldFormat[article]{volume}{\mkbibbold{#1}}
\DeclareFieldFormat[article]{number}{\mkbibparens{#1}}
\DeclareFieldFormat[article]{title}{\MakeCapital{#1}}
\DeclareFieldFormat[inproceedings]{title}{#1}
\DeclareFieldFormat{shorthandwidth}{#1}
\usepackage{xpatch}
\xpatchbibmacro{volume+number+eid}{\setunit*{\adddot}}{}{}{}
\renewbibmacro{in:}{%
  \ifentrytype{article}{}{%
    \printtext{\bibstring{in}\intitlepunct}}}

\makeatletter
\DeclareDelimFormat[cbx@textcite]{nameyeardelim}{\addspace}
\makeatother

\RequirePackage[absolute,overlay]{textpos}
\date{}

\usepackage{tikz}
\usepackage{algorithm}
\usepackage{algpseudocode}
\usepackage{amsthm}
\usepackage{amsmath,bm}
\usepackage{paralist}
\usepackage{todonotes}
\usepackage{ctable}
\usepackage{multirow}
\usepackage{lscape}
\usepackage{rotating}
\usepackage{float}
\usepackage{enumitem}
\usepackage[referable]{threeparttablex}

\allowdisplaybreaks
\def\yes{$\checkmark$}

\begin{document}
\maketitle
\begin{abstract}
  \noindent This paper introduces a novel meta-learning algorithm for time series forecast model performance prediction. We model the forecast error as a function of time series features calculated from the historical time series with an efficient Bayesian multivariate surface regression approach. The minimum predicted forecast error is then used to identify an individual model or a combination of models to produce the final forecasts. It is well-known that the performance of most meta-learning models  depends on the representativeness of the reference dataset used for training. In such circumstances, we augment the reference dataset with a feature-based time series simulation approach, namely GRATIS, in generating a rich and representative time series collection. The proposed framework is tested using the M4 competition data and is compared against commonly used forecasting approaches. Our approach provides comparable performances to other model selection/combination approaches but at a lower computational cost and a higher degree of interpretability, which is important for supporting decisions. We also provide useful insights regarding which forecasting models are expected to work better for particular types of time series, the intrinsic mechanisms of the meta-learners and how the forecasting performances are affected by various factors.

\end{abstract}

\newpage
\hypertarget{intro}{%
  \section{Introduction}\label{intro}}

Forecasting is an important aspect of every business operation. The selection of a
suitable forecast model or a combination of models to use in forecasting is at the heart
of the forecasting process \autocite{tashman1991automatic}. However, this
selection/combination process is challenging in the context of large-scale time series
forecasting for the reasons that : (i) there is no universal method that performs best for
all kinds of forecasting problems, (ii) a trial-and-error process of model selection would
increase the computational time as well as the computational cost significantly, (iii) it
is not possible to derive a typical algebraic expression for choosing the best-performing
model(s) out of a portfolio of algorithms, and (iv) the traditional expert's judgement
cannot be quickly scaled to forecast a large number of series problems due to the
computational cost and time constraints. Therefore, automatic forecast model selection is
essential and challenging in nowadays business. Automatic forecast model selection
approaches often test multiple forecasting algorithms and select the most appropriate one
according to some criterion, such as information criterion~\citep{Hyndman2008} and
out-of-sample tests~\citep{tashman2000out}. Meta-learning serves as a promising
alternative to solve this problem.  \citet{petropoulos2018exploring} state that there are
model, parameter and data uncertainty in forecasting. That means that even if the
above-mentioned challenges could be effectively addressed, there would still be different
forms of uncertainties, which meta-learning could help in mitigating by learning across
different time series and datasets.

The idea of using meta-learning to select the best forecasting model for a given time
series has been explored by many researchers in the forecasting community
\autocites{collopy1992rule}{shah1997model}{adya2001automatic}{wang2009rule}{Petropoulos2014}. This
approach is also known as an algorithm selection problem and can be expressed firmly using
Rice's framework for algorithm selection \autocite{rice1976}. More recently,
~\citet{Petropoulos2014} has pointed out that there are \textit{horses for courses}, which
also supports the idea of meta-learning. Further evidence in favour of this idea is also
given in \textcite{fforms}. In all these cases, a vector of features computed from time
series is used as an input to train a meta-learner. The output and the objective function
used in the meta-learner are approached differently. For example, \textcite{shah1997model}
uses the best forecast model as the output label and applies discriminant analysis to
predict the forecast model that is expected to perform best on a given time series and the
objective function is to minimise classification error. Also,
\textcite{prudencio2004using} use neural network approaches to identify weights for the
best linear combination of methods to improve forecast accuracy; thus, the objective
function is to minimise the forecast error. \citet{Petropoulos2014} translate seven time
series features and the forecasting horizon into forecasting model selection making the
forecasting horizon directly linked to the short/long-term forecast tasks.

Although many researchers have highlighted the usefulness of the meta-learning approach to compute the forecasts, few studies could conclude that its approach is constantly superior to simple benchmarks and commonly used forecasting approaches. For example, \textcite{meade2000evidence} depicts that the summary statistics are useful in selecting a suitable forecasting method, but are not necessarily the best. Three possible reasons for the infeasibility of the selection process are the use of inadequate features, the improper choice of meta-learning algorithms, and having training time series data that are not as diverse as required to predict different forecast model performance \autocite{kang2020gratis}. Recently, several studies strive towards these directions. For example, the results of recent large-scale M4-competition highlight features-based algorithms offer a promising solution for time series forecasting. \citet{Li2020} use time series imaging to extract a comprehensive set of time series features as the input of the meta-learning framework. \citet{wang2021uncertainty}, for the first time, investigate feature-based interval forecasting. Therefore, developing new or improving existing feature-based forecasting approaches might be a worthwhile endeavour.

We address the aforementioned issues in developing a meta-learning framework for forecast model selection based on features computed from the time series. Our first attempt to develop a framework for forecast model selection is described in \textcite{fforms}. The first framework is called \textsf{FFORMS}: Feature-based FORecast Model-Selection, in which we use a classification algorithm (random forest) to predict the ``best forecast model'' on a given time series. Our following-up essay, \textcite{fforma}, rather than mapping time series to a single forecast model, uses a gradient boosting algorithm to obtain the weights for forecast combinations. The second framework is named \textsf{FFORMA}: Feature-based FORecast Model Averaging. \textsf{FFORMA} placed second in the M4 competition \autocite{makridakis2018m4}. This paper, which is the third in this series, considered the correlation structure of algorithm performance in their model training process.  Having revisited the literature, we found that, to the best of our knowledge, few of these studies have tackled the forecast model performance prediction with meta-learning. In a conventional univariate supervised meta-learning problem, merely labelling the instances using the ``best forecast model'' is difficult for the meta-learner to correctly retrieve the interconnections between individual forecast models. The current paper further extends the idea of meta-learning by making the dependent variables a multivariate forecast errors vector. Hence, the meta-model is trained to learn the interrelationship between the forecast errors across different class labels.

\begin{figure}
\centering
\includegraphics[width=0.97\linewidth]{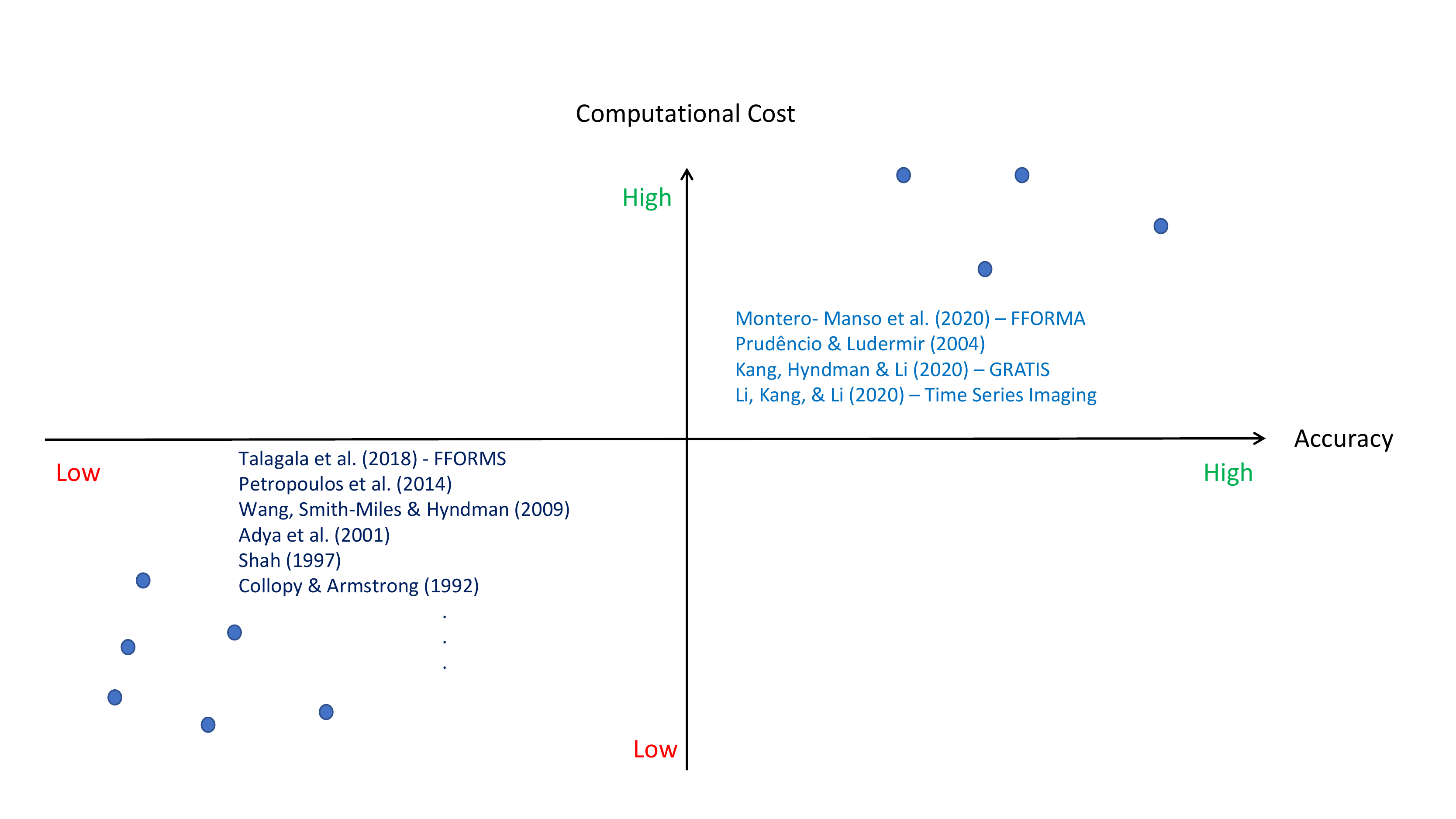}
  \caption{Classification of existing meta-learning algorithms based on computational cost and accuracy.}
  \label{fig:complexity}
\end{figure}

The existing meta-learning algorithms for forecasting are divided into two categories: (1)
selecting the ``best forecast model'' (the model with the smallest forecast error) out of
a pool of forecast models, and (2) identifying suitable weights to combine forecasts from
all available models in the pool. Based on this, the meta-learning algorithms can be
further classified into two extremes in terms of forecasting accuracy and computational
cost: (1) low computational cost and low accuracy, and (2) high computational cost and
high accuracy. There is a continuum between the two extremes. All existing approaches fall
at one extreme of the spectrum: (1) low-low: forecasting based on a single best model, and
(2) high-high: forecasting based on all available models in the pool.  In this paper, we
treat the forecast model selection problem as a ranking problem. We propose an algorithm
to rank forecast models by simultaneously predicting the forecast errors. The rankings
allow the user to identify a subset of forecasting models. This way, we can gain a good
balance between the accuracy and computational cost, which allows the algorithm to move
between the two extremes of the spectrum. Therefore, we refer to our new framework as
\textsf{FFORMPP}: Feature-based FORecast Model Performance Prediction. A schematic
illustration is given in \autoref{fig:complexity}. \autoref{metalit} also provides an
overview of some of the meta-learners available in the literature.

The proposed algorithm is closely related to the concept of forecast pooling, i.e., selecting and combining from a pool of suitable models instead of all models available in the model. For example, \citet{aiolfi2006persistence} propose a conditional forecast combination approach that first cluster the candidate models, pools forecasts within each cluster and estimates their optimal weights. \citet{matsypura2018optimal} combine expert forecasts by forecast pooling. \textcite{kourentzes2019another} introduces the forecast island approach for forecast selection and combination. The authors investigate the use of quantiles to form groups and proposed a heuristic to automatically identify the forecast pool rather than using a model to decide which forecasts should be included in the combination or not.

\begin{sidewaystable}
  \centering
  \caption{An overview of literature for time series forecasting with meta-learning
    approaches.}
  \label{metalit}
\resizebox{30cm}{7cm}{
    \begin{tabular}{p{6cm}p{4cm}p{3cm}lp{6cm}p{2.5cm}p{3cm}}
      \toprule
      Authors (Year)                        & Type of meta-learner (selection, combination, both) & Output                              & Type of forecasts & Meta-learning algorithm                                                          & Number of features used                    & Number of forecasting methods             \\
      \midrule
      \citet*{arinze1997combining}          & both                                                & error                               & point               & rule based induction                                                             & 6                                          & 6                                         \\
      \citet*{shah1997model}                & selection                                           & label                               & point               & discriminant analysis                                                            & 26                                         & 3                                         \\
      \citet*{venkatachalam1999intelligent} & selection                                           & error                               & point               & two-stage neural network                                                         & 6                                          & 9                                         \\
      \citet*{adya2000application}          & both                                                & error                               & point               & rule based (judegemental)                                                        & 18                                         & 3                                         \\
      \citet*{prudencio2004meta}: study 1   & both                                                & error                               & point               & C4.5 decision tree  in \citep{hall2009weka}                                                              & 14                                         & 2                                         \\
      \citet*{prudencio2004meta}: study 2   & both                                                & ranking based on error              & point               & neural network classifier                                                        & 16                                         & 3                                         \\
      \citet*{wang2006characteristic}       & selection                                           & error                               & point               & C4.5 decision tree                                                               & 13                                         & 4                                         \\
      \citet*{wang2009rule}                 & selection                                           & error                               & point               & SOM clustering and DT algorithm                                                  & 9                                          & 4                                         \\
      \citet*{lemke2010meta}                & both                                                & error                               & point               & i) a feed forward neural network, ii) decision tree, iii) support vector machine & 24                                         & 15                                        \\
      \citet*{widodo2013model}              & selection                                           & label                               & point               & K-Nearest Neighbor                                                               & 13 features by \citet{wang2009rule}        & 6                                         \\
      \citet*{Petropoulos2014}              & selection                                           & error                               & point               & multiple regression                                                              & 14                                         & 7                                         \\
      \citet*{kuck2016meta}                 & selection                                           & label                               & point               & neural network                                                                   & 127                                        & 4                                         \\
      \citet*{fforms}                       & selection                                           & label                               & point               & random forest algorithm                                                          & yearly - 26 , quarterly - 30, monthly - 30 & yearly - 9 , quarterly - 14, monthly - 14 \\
      \citet*{fforma}                       & combination                                         & weights for forecasting combination & point               & XGBoost                                                                          & 42                                         & 9                                         \\
      \citet*{Li2020}                       & combination                                         & error                               & point               & XGBoost                                                                          & time series imaging features               & 9                                         \\
      \citet*{wang2021uncertainty}          & both                                                & error                               & interval            & GAM models                                                                       & 40                                         & 8                                         \\
      \bottomrule
    \end{tabular}
  }
\end{sidewaystable}

In our approach, we utilise the multivariate structure in the response variables to simultaneously model the correlation structure between different forecast model errors using the time series features. The Bayesian multivariate surface regression approach proposed by \textcite{li2013efficient} is used to estimate the forecast error for each model in the pool. This allows the ranking of the forecast models with respect to their forecast errors and the evaluation of their relative forecast performance without calculating forecasts from all available individual models in the pool. Our framework merits the following contributions to the forecasting domain.

\begin{enumerate} \def\labelenumi{\arabic{enumi}.}
\item The diversity of features in the collection of time series used to train a meta-learner plays a critical role in training a meta-learning model. Often, time series with the required amount of feature diversity and quality might not be available \autocites{kang2017visualising}{kang2020gratis}. To this end, we augment the original reference dataset with \textsf{GRATIS} proposed by \textcite{kang2020gratis} to obtain a diverse collection of time series. We explore whether enriching the diversity of the training set helps in increasing the accuracy of forecasts.

\item One noticeable gap in the meta-learning literature is that little attempt has been made to identify the relationships learned by the meta-learning algorithm. To this end, we visualise the relationships learned by the meta-learning algorithm. The visualisation approach involves mapping each time series as a point in a two-dimensional instance space given by the features and exploring the relationship between features and forecast models selected. This helps to gain insights into why specific models perform better on certain types of time series, which is also useful in the understanding of the strengths and weaknesses of the meta-learning algorithm.

\item We utilise an efficient Bayesian multivariate surface regression model to train the
  meta-learner. Compared with other meta-learning algorithms, our approach treats the
  features as covariates in the regression form for improving the meta-learning
  interpretability, because linear and nonlinear effects can be separated with stochastic knots. Unlike the univariate
  regression models that do not take into account the relationships between features and
  their correlations to the classification between forecasting models, The multivariate
  surface regression model jointly models the relationship between forecast errors and
  time series features for multiple forecasting models.
\end{enumerate}


The remainder of the paper is organised as follows: \autoref{methodology} introduces the methodology, including the methodology for simulating time series to augment the reference set to train a model and efficient Bayesian multivariate regression approach. \autoref{results} discusses the results in application to the M4 competition data. \autoref{conclusion} concludes the paper.

\hypertarget{methodology}{%
  \section{Methodology}\label{methodology}}

\begin{figure}
\centering \includegraphics[width=\linewidth]{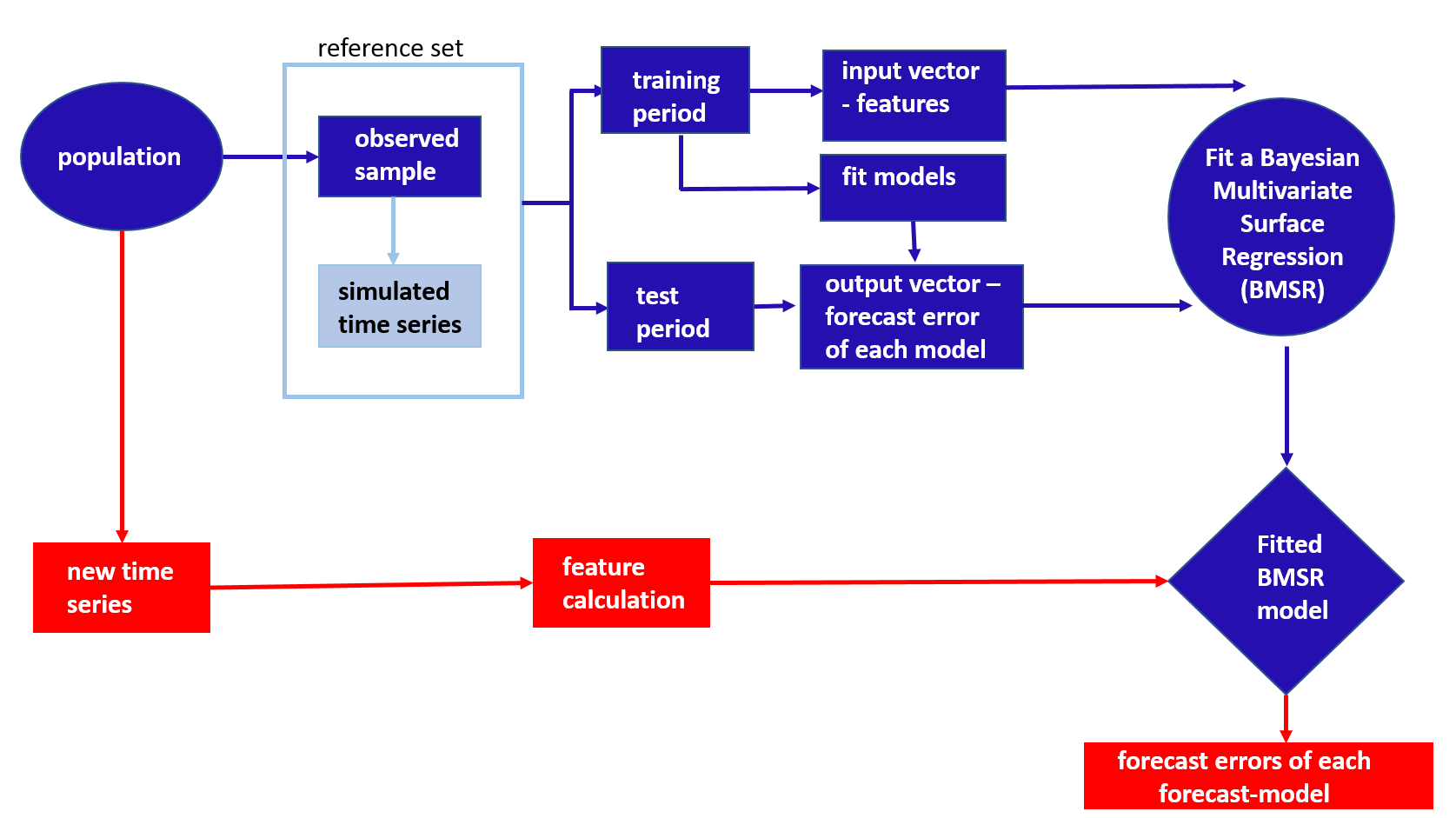}
  \caption{\textsf{FFORMPP} (Feature-based FORecast Model Performance Prediction) framework. The offline phase is shown in blue and the online phase in red.}
  \label{fig:framework}
\end{figure}

Our \textsf{FFORMPP} framework (\autoref{fig:framework}) consists of two phases: the offline phase and the online phase.  We treat the forecast model selection problem as a multi-class ranking problem, where predictors are features, and the outcome consists of forecast errors (MASE) calculated over all available models. In particular, an instance \(i\) is a tuple \((f_{1i}, f_{2i}, \cdots , f_{mi},e_{1i}, e_{2i}, \cdots , e_{ni})\), where predictors \(( f_{1i}, f_{2i}, \cdots, f_{mi})\) is a vector of \(m\) features which is the input to our algorithm, and the outcome \((e_{1i}, e_{2i}, \cdots , e_{ni})\), is a vector of MASE values of the all \(n\) forecast models. A description of offline and online phases are as follows.

\begin{itemize}
\item \emph{Offline phase:}

  \begin{enumerate}
  \item \textbf{Reference set and its augmentation with \textsf{GRATIS}. } We take an observed sample of time series from the population of interest. In addition to the observed time series, we augment the reference set by adding simulated time series. This data simulation part is an optional component in our framework. This is needed only if the practitioners have a limited observed sample. If the observed sample is large and diverse enough, there is no need to add simulated time series.

  \item \textbf{Training and testing periods splitting}. We split all time series in the reference set into a training period and a test period once we created the reference set. For each series in the reference set, we compute features based on the training period. The calculated features become the input vector to our meta-learner.

  \item \textbf{Time series features and model class labels}. The output vector for each series consists of Mean Absolute Scaled Error (MASE) of the all available candidate models. To compute the MASE, all candidate models are estimated using the training period, and forecasts are generated for the whole of the test period (based on "a fixed origin" evaluation). Subsequently, MASE for each model is computed over the test period. This completes the data processing part of our algorithm.

  \item \textbf{Meta-learner training}. Finally, an efficient Bayesian multivariate surface regression (EBMSR) approach proposed by \textcite{li2013efficient} is used to model forecast error measures (which in our case is MASE) as a function of features calculated from the time series. This produces a meta-learner to be used in the online phase.

\end{enumerate}

\item \emph{Online phase:} The online phase requires only the calculation of a simple vector of features for any newly given time series. It uses the pre-trained classifier to estimate forecast error for each candidate model in the pool. Finally, forecast models are ranked based on the predicted errors. The top ranking models are used to produce forecasts.

\end{itemize}

Most of the expensive computations for processing data and training a meta-learner are performed in the offline phase. In the following sections, each component in the offline phase is explained in detail with application to forecasting M4 competition data. In our application, we treat M4 competition data as the new time series to be forecast in the online phase.

\subsection{Reference set and its augmentation with \textsf{GRATIS}}\label{reference-set}

We use the time series of the M1 and M3 competitions as the observed sample for yearly, quarterly and monthly time series. We also simulate 10, 000 time series with \textsf{GRATIS}\footnote{The R package \texttt{gratis} accompanies this work and is publicly available on \url{https://CRAN.R-project.org/package=gratis}.}~\citep{kang2020gratis} to augment the reference set. The observed time series and the simulated time series form the reference set to fit the model. For weekly, daily and hourly frequencies, only the simulated time series are used to create the reference set because the previous M-competitions do not contain the time series corresponding to weekly, daily and hourly frequencies.

The augmentation with simulated time series helps obtain a more heterogeneous collection of time series for training a meta-learner, which helps prevent overfitting to a relatively homogeneous set of data and increases generalisability of the model when applied to new time series with different conditions. Furthermore, the augmentation is particularly beneficial when there is no observed set of time series available to train a model in the offline phase in e.g. judgemental forecast, business forecasting with new products.
Although there is no standard process for simulating time series, the most common approach involves simulation based on some data-generating processes (DGPs) such as exponential smoothing and ARIMA models. Instead of relying on a set of DGPs to generate time series, the \textsf{GRATIS} algorithm simulates time series with diverse time series features using mixture autoregressive (MAR) models. The algorithm can also allow a user to set controllable features for simulating time series.

\autoref{marparameters} lists the choice of values for parameters used to simulate time series in each frequency category. We used frequencies 1, 4, 12 and 52 to generate yearly, quarterly, monthly, and weekly series, respectively. Daily and hourly time series with a long history often show multiple seasonal patterns. Hence, for daily series, frequencies were set to 7 (time-of-week pattern) and 365.25 (annual seasonality), while for hourly series frequencies were set to 24 (time-of-day pattern) and 168 (time-of-week pattern). None of the time series in our hourly test dataset is longer than 8760. Hence, time-of-year pattern (\(365\times24=8760\)) was not considered. Except for the hourly series, the length of the time series is randomly chosen from the uniform distribution. The minimum and maximum values of the distributions are selected based on lower (Q1 - 1.5 \(\times\) IQR) and upper (Q3 + 1.5 \(\times\) IQR) edges of the box-and-whisker plots.

\begin{figure}
  \centering
  \includegraphics[width=\textwidth]{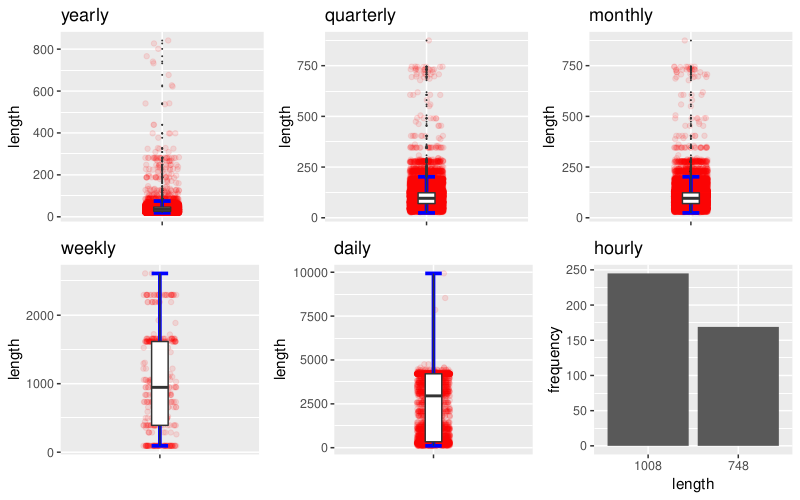}
  \caption{Distribution of the time series lengths of M4 across different frequency categories. The data points are shown in red. The horizontal blue solid line shows the lower and the upper whisker borders of each boxplot. Some outliers could be observed in yearly, quarterly and monthly categories. There are only two distinct levels of length for hourly series.}
  \label{fig:lengthboxplot}
\end{figure}

We use the M4 competition data to compute the associated statistics for length. The
corresponding distributions are shown in \autoref{fig:lengthboxplot}. The reason for the
aforementioned choice is that we use the length of the time series as a feature in our
meta-learning framework. Hence, the lengths of the series in our reference set should cover
all or the majority of time series we need to forecast. In practice, it might be difficult
to obtain a reference set covering the whole length of the new time series for two main
reasons: (i) information regarding the range of length is not available at the offline
stage (although, a rough idea about the distribution of the majority can be obtained), and
(ii) the range of length is very wide owing to `outlying' observations (for example,
\autoref{fig:lengthboxplot}, quarterly and monthly series). In such circumstances, a
reference set covering most of the lengths of future time series is a reasonable
approach. Further, for each series we randomly select a number of mixing components from
\(\{1, 2, 3, 4, 5\}\). This is because \citet{li2010flexible} and
\citet{villani2009regression} point that, for mixture models with comprehensive mean
structures, at most five components are sufficient in order to capture widely shapes of the data.  Other parameters are set with random values from certain
distributions as used in \textsf{GRATIS}. These are analogous to non-informative priors in the
Bayesian contexts, i.e., the diversity of the generated time series should not rely on the
parameter settings. Having each parameter set as in \autoref{marparameters}, we generate
10000 time series from each frequency category. This is to reduce the training time of the
efficient Bayesian multivariate surface regression model while maintaining the diversity
of the reference set.

As shown in Figure 4, the use of simulated data generated by \textsf{GRATIS} along with M1 and M3 competition data allows us to examine if the simulated data generated based on the \textsf{GRATIS} approach are actually representative of real data. Note that training the algorithms only using simulated time series could be dangerous in the case the simulated series may not be a representative sample of the population of interest. However, if the user does not have an option to obtain the required data, then the only option is to use simulated data. In such circumstances, extra care needs to be taken when setting parameter values of the \textsf{GRATIS} algorithm.

\begin{table}
  \caption{Choice of values for parameters for simulating time series using \textsf{GRATIS}.}
  \label{marparameters}
  \resizebox{\textwidth}{!}{
  \begin{tabular}{llp{2cm}p{4cm}}
    \toprule
    Parameter                                 & Description                                      & \multicolumn{2}{p{2cm}}{Value}                                                             \\
 \midrule \multirow{6}{*}{\texttt{frequency}} & \multirow{6}{*}{seasonal period}                 & yearly    & 1                                                                              \\
                                              &                                                  & quarterly & 4                                                                              \\
                                              &                                                  & monthly   & 12                                                                             \\
                                              &                                                  & weekly    & 52                                                                             \\
                                              &                                                  & daily     & (7, 365)                                                                       \\
                                              &                                                  & hourly    & (24, 168)                                                                      \\
 \midrule \multirow{6}{*}{\texttt{length}}    & \multirow{6}{*}{length of time series}           & yearly    & U(19, 75)                                                                      \\
                                              &                                                  & quarterly & U(24, 202)                                                                     \\
                                              &                                                  & monthly   & U(60, 660)                                                                     \\
                                              &                                                  & weekly    & U(93, 2610)                                                                    \\
                                              &                                                  & daily     & U(107, 9933)                                                                   \\
                                              &                                                  & hourly    & 40.8\% with 748 length and 59.2\% with 1008 length                             \\
    \midrule
    k                                         & number of mixing components                      & \multicolumn{2}{p{5cm}}{For each series randomly chosen from $\{1, 2, 3, 4,5\}$}           \\
 $\alpha_k$                                   & weights of mixture components                    & \multicolumn{2}{l}{$\alpha_k=\beta_k/\sum_{i=1}^{K}\beta_i$, where $\beta_i \sim U(0, 1)$} \\
 $\theta_{ki}$                                & coefficients of the AR part                      & \multicolumn{2}{l}{N(0, 0.5)}                                                              \\
 $\Theta_{kj}$                                & coefficients of the seasonal AR parts            & \multicolumn{2}{l}{N(0, 0.5)}                                                              \\
 $d_k$                                        & number of differences in each component          & \multicolumn{2}{l}{Bernoulli(0.9)}                                                         \\
 $D_k$                                        & number of seasonal differences in each component & \multicolumn{2}{l}{Bernoulli(0.4)}                                                         \\
 \bottomrule
  \end{tabular}
  }
\end{table}

\subsection{Training and test periods splitting}\label{training-period-and-test-period}

Once we have formed the reference set, we split each time series in the reference set into training period and test period. Features are calculated based on the training period of each series, and the forecast error measure is calculated over the test period of each time series. \autoref{observedsample} summarises the number of time series in the reference set and the new time series in each frequency category. Note that for yearly, quarterly and monthly time series, the reference set used to train a meta-learner is much smaller than the test set evaluating the meta-learner. This is often the case when applying the meta-learner in practice during the online phase.

\begin{table}
\centering
  \caption{Composition of the time series in the reference set and collection of new time series.}
  \label{observedsample}
  \begin{tabular}{lrrrr}
    \toprule
    \multirow{2}{*}{Frequency} & \multicolumn{3}{c}{Reference set} & New series               \\
 \cline{2-4}                   & M1                                & M3   & Augmented & M4    \\
 \midrule Yearly               & 181                               & 645  & 10000     & 23000 \\
 Quarterly                     & 203                               & 756  & 10000     & 24000 \\
 Monthly                       & 617                               & 1428 & 10000     & 48000 \\
 Weekly                        & -                                 & -    & 10000     & 359   \\
 Daily                         & -                                 & -    & 10000     & 4227  \\
 Hourly                        & -                                 & -    & 10000     & 414   \\
 \bottomrule
  \end{tabular}
\end{table}

\subsection{Time series features and model class labels}

For each series, a set of features computed from the training period of each series comprising an input vector and the Mean Absolute Scaled Error~\citep[MASE, ][]{hyndman2006another} values for each method give the output vector to train a model. In this paper, we use MASE as a measure of out-of-sample forecasting accuracy because it is less dependent on the data scales. Hence, MASE can be used to compare forecasts across time series with different scales. Furthermore, \textcite{hyndman2006another} describe MASE as a ``generally applicable measurement of forecasting accuracy without the problems seen in the other measurements.'' Besides, MASE is one of the main measurements of forecasting accuracy used in forecasting competitions. For a time series $y_t,~t = 1, 2, \cdots, T$, MASE is defined by
\begin{equation}
\text{MASE}=\frac{\frac{1}{H}\sum_{h=1}^H |y_{T+h}-\hat{y}_{T+h}|}{\frac{1}{n-m}\sum_{t={m+1}}^{T}|y_t - y_{t-m}|},
\end{equation}
where \(H\) is the length of the forecast horizon, \(T\) is the length of the training period of the time series and \(m\) is the frequency of the time series. $\hat{y}_{T+h}$ is the $h$-step point forecast. The description of the features calculated in each frequency category is shown in \autoref{feature}. We analyse yearly, quarterly, monthly, weekly, daily and hourly series separately. \autoref{classlabels} shows the forecast models we consider within each frequency category.

\begin{table}
  \caption{Time series features and availabilities for yearly (Y), quarterly (Q), monthly (M), weekly (W), daily (D) and hourly (H) data. }
  \label{feature}
  \begin{tabular}{llp{8cm}cccc}
    \toprule
    Index & Feature                 & Description                                                                             & Y     & Q/M  & W    & D/H  \\
    \midrule
1         & \texttt{T}              & length of time series                                                                   & \yes  & \yes & \yes & \yes \\
 2        & \texttt{trend}          & strength of trend                                                                       & \yes  & \yes & \yes & \yes \\
 3        & \texttt{seasonality\_q} & strength of quarterly seasonality                                                       & -     & \yes & -    & -    \\
 4        & \texttt{seasonality\_m} & strength of monthly seasonality                                                         & -     & \yes & -    & -    \\
 5        & \texttt{seasonality\_w} & strength of weekly seasonality                                                          & -     & -    & \yes & \yes \\
 6        & \texttt{seasonality\_d} & strength of daily seasonality                                                           & -     & -    & -    & \yes \\
 7        & \texttt{seasonality\_y} & strength of yearly seasonality                                                          & -     & -    & -    & \yes \\
 8        & \texttt{linearity}      & linearity                                                                               & \yes  & \yes & \yes & \yes \\
 9        & \texttt{curvature}      & curvature                                                                               & \yes  & \yes & \yes & \yes \\
 10       & \texttt{spikiness}      & spikiness                                                                               & \yes  & \yes & \yes & \yes \\
 11       & \texttt{e\_acf1}        & first ACF value of remainder series                                                     & \yes  & \yes & \yes & \yes \\
 12       & \texttt{stability}      & stability                                                                               & \yes  & \yes & \yes & \yes \\
 13       & \texttt{lumpiness}      & lumpiness                                                                               & \yes  & \yes & \yes & \yes \\
 14       & \texttt{entropy}        & spectral entropy                                                                        & \yes  & \yes & \yes & \yes \\
 15       & \texttt{hurst}          & Hurst exponent                                                                          & \yes  & \yes & \yes & \yes \\
 16       & \texttt{nonlinearity}   & nonlinearity                                                                            & \yes\ & \yes & \yes & \yes \\
 17       & \texttt{alpha}          & ETS(A,A,N) $\hat\alpha$                                                                 & \yes  & \yes & \yes & -    \\
 18       & \texttt{beta}           & ETS(A,A,N) $\hat\beta$                                                                  & \yes  & \yes & \yes & -    \\
 19       & \texttt{hwalpha}        & ETS(A,A,A) $\hat\alpha$                                                                 & -     & \yes & -    & -    \\
 20       & \texttt{hwbeta}         & ETS(A,A,A) $\hat\beta$                                                                  & -     & \yes & -    & -    \\
 21       & \texttt{hwgamma}        & ETS(A,A,A) $\hat\gamma$                                                                 & -     & \yes & -    & -    \\
 22       & \texttt{ur\_pp}         & test statistic based on Phillips-Perron test                                            & \yes  & -    & -    & -    \\
 23       & \texttt{ur\_kpss}       & test statistic based on KPSS test                                                       & \yes  & -    & -    & -    \\
 24       & \texttt{y\_acf1}        & first ACF value of the original series                                                  & \yes  & \yes & \yes & \yes \\
 25       & \texttt{diff1y\_acf1}   & first ACF value of the differenced series                                               & \yes  & \yes & \yes & \yes \\
 26       & \texttt{diff2y\_acf1}   & first ACF value of the twice-differenced series                                         & \yes  & \yes & \yes & \yes \\
 27       & \texttt{y\_acf5}        & sum of squares of first 5 ACF values of original series                                 & \yes  & \yes & \yes & \yes \\
 28       & \texttt{diff1y\_acf5}   & sum of squares of first 5 ACF values of differenced series                              & \yes  & \yes & \yes & \yes \\
 29       & \texttt{diff2y\_acf5}   & sum of squares of first 5 ACF values of twice-differenced series                        & \yes  & \yes & \yes & \yes \\
 30       & \texttt{sediff\_acf1}   & ACF value at the first lag of seasonally-differenced series                             & -     & \yes & \yes & \yes \\
 31       & \texttt{sediff\_seacf1} & ACF value at the first seasonal lag of seasonally-differenced series                    & -     & \yes & \yes & \yes \\
 32       & \texttt{sediff\_acf5}   & sum of squares of first 5 autocorrelation coefficients of seasonally-differenced series & -     & \yes & \yes & \yes \\
 33       & \texttt{seas\_pacf}     & partial autocorrelation coefficient at first seasonal lag                               & -     & \yes & \yes & \yes \\
 34       & \texttt{lmres\_acf1}    & first ACF value of residual series of linear trend model                                & \yes  & -    & -    & -    \\
 35       & \texttt{y\_pacf5}       & sum of squares of first 5 PACF values of original series                                & \yes  & \yes & \yes & \yes \\
 36       & \texttt{diff1y\_pacf5}  & sum of squares of first 5 PACF values of differenced series                             & \yes  & \yes & \yes & \yes \\
 37       & \texttt{diff2y\_pacf5}  & sum of squares of first 5 PACF values of twice-differenced series                       & \yes  & \yes & \yes & \yes \\
 \bottomrule
  \end{tabular}
\end{table}

\begin{table}
  \caption{Forecasting model class label and availabilities for yearly (Y), quarterly (Q), monthly (M), weekly (W), daily (D) and hourly (H) data.}
  \label{classlabels}
  \begin{tabular}{lp{0.5\columnwidth}cccc}
    \toprule
    Model class label & Description                                                                                                                                                                           & Y          & Q/M        & W          & D/H        \\
    \midrule
    \textsf{WN}       & White noise process                                                                                                                                                                   & \checkmark & \checkmark & \checkmark & \checkmark \\
 \textsf{auto.arima}  & The autoregressive integrated moving average model with automatic lag selections.                                                                                                     & \checkmark & \checkmark & \checkmark & -          \\
 \textsf{ets}         & The exponential smoothing state space model.                                                                                                                                          & \checkmark & \checkmark & -          & -          \\
 \textsf{rw}          & Random walk                                                                                                                                                                           & \checkmark & \checkmark & \checkmark & \checkmark \\
 \textsf{rwd}         & Random walk with drift.                                                                                                                                                               & \checkmark & \checkmark & \checkmark & \checkmark \\
 \textsf{theta}       & The decomposition forecasting model by modifying the local curvature of the time-series through a coefficient `Theta' that is applied directly to the second differences of the data. & \checkmark & \checkmark & \checkmark & \checkmark \\
 \textsf{stlar}       & The STL decomposition with AR modeling of the seasonally adjusted series.                                                                                                             & -          & \checkmark & \checkmark & \checkmark \\
 \textsf{snaive}      & The seasonal na\"ive method, which forecasts using the most recent values of the same season.                                                                                         & -          & \checkmark & \checkmark & \checkmark \\
 \textsf{tbats}       & The exponential smoothing state space model with a Box-Cox transformation, ARMA errors, trend and seasonal components                                                                 & -          & \checkmark & \checkmark & \checkmark \\
 \textsf{nn}          & Neural network time series forecasts.                                                                                                                                                 & \checkmark & \checkmark & \checkmark & \checkmark \\
 \textsf{mstlets}     & Multiple seasonal time series forecasts with ETS.                                                                                                                                     & -          & -          & \checkmark & \checkmark \\
 \textsf{mstlarima}   & Multiple seasonal time series forecasts with ARIMA.                                                                                                                                   & -          & -          & -          & \checkmark \\
 \bottomrule
  \end{tabular}
\end{table}

\subsection{Training the meta-learner}

We use a Bayesian surface spline regression to train the meta-learner. The most commonly used additive spline regression assumes additivity in the regressor; that is, \(E(y|x_1, ..., x_q)=\sum_{j=1}^{q}f_j(x_j),\) where \(f_j(x_j)\) is a spline regressor of the \(j^{th}\) regressor. Even though the assumption of additivity simplifies the model, it is quite a restrictive assumption. This problem has motivated research on surface models with interactions between regressors \autocite{li2013efficient}. \textcite{li2013efficient} proposed a general Bayesian approach for fitting surface models for a continuous multivariate response by combining additive splines and interactive splines. The proposed modelling is called efficient Bayesian multivariate surface regression.

The reason for using multivariate analysis was that univariate analysis does not take into account the relationships between variables and their correlations to the classification between groups. The motivation for the use of efficient Bayesian multivariate surface regression is due to its flexibility. The main challenge in spline regression is the choice of knot locations.

The number of knots has an important influence on the resulting fit of spline regression: without enough knots, the regression is underfitted, and with too many knots, it is overfitted. Choosing the locations of knots is also a challenge. This becomes even harder for surface regression than it is for additive models because any feasible set of \(q\)-dimensional knots is necessarily sparse in \(R^q\) when the number of regressors, \(q\), is moderate or large. This causes the curse of dimensionality. The most common approach used in the literature is to use a fixed set of knot locations, and most of these algorithms place the knots at the centroids of the clusters computed based on regressor observations. \textcite{li2013efficient} pointed out that this is impractical when estimating a surface with several regressors. Hence, the authors use a computationally efficient Markov chain Monte Carlo (MCMC) algorithm for the Gaussian multivariate surface regression to update the locations of the knots jointly. Instead of a fixed set of locations, the authors introduce \textsf{moving knots} that allows the location of the knots to move freely in the regressor space.

The proposed Gaussian multivariate regression model can be written as follows:
\begin{equation}
  \label{eq:1} \boldsymbol{Y=X_0B_0+{X_a}(\xi_a)B_a+{X_s}(\xi_s)B_s+E},
\end{equation} where \(\bold{Y}\) is a matrix of \(n\) number of observations and \(p\) number of response variables. The rows of \(\bold{E}\) are error vectors assumed to be independent and identically distributed (iid) as \(N_p(0,\Sigma)\). The proposed efficient Bayesian multivariate surface regression model contains three components:

\begin{enumerate}
\item \textbf{Linear component: } The linear component contains the original covariates including the constant term. This enters the model in linear form. The matrix \(\bold{X_0}\) is an \(n \times q_0\) vector in which the first column contains ones for the intercept. The corresponding regression coefficients are in \(\bold{B_o}\).

\item \textbf{Additive component: } The second component of the model contains additive spline basis functions of the covariates in \(\bold{X_0}\). This is represented by \(\boldsymbol{X_a(\xi_a)}\), where \(\xi_a\) represents the knots. It is important to note that the knots in the additive part of the model are scalars and this model allows an unequal number of knots for different covariates in the model. The matrix \(\bold{B_a}\) contains the regression coefficients corresponding to the additive component. The additive component of the model captures the nonlinear relationship between features and response \(\bold{Y}\).

\item \textbf{Surface component: } The surface component of the model contains the radial basis function for capturing the remaining part of the surface and interactions. This is denoted by \(\boldsymbol{X_s(\xi_s)}\). Note that the \(\xi_s\) is a \(q_0\)-dimensional vector. The matrix \(\bold{B_s}\) contains the regression coefficients corresponding to the surface component and \(\xi_s\) represents the surface knots.
\end{enumerate}

For notational convenience Equation \eqref{eq:1} can be written as
\begin{equation} \mathbf{Y}=\mathbf{XB+E},
  \label{eq:eqn2}
\end{equation} where \(\bold{X=[X_0, X_a, X_s]}\) is the \(n \times q\) design matrix (\(q=q_0+q_a+q_s\)) and \(\bold{B=[B_0^{'}, B_a^{'}, B_s^{'}]}\). For a given set of fixed knot locations, the model in Equation \eqref{eq:1} is linear in regression parameters. The over-parameterised problem is addressed by using shrinkage priors to shrink small regression coefficients towards zero. The shrinkage parameters and knot locations of \(\xi_a\) and \(\xi_s\) are treated as unknown parameters to be estimated. \textcite{li2013efficient} proposed a computationally efficient MCMC algorithm to estimate shrinkage parameters and update the locations of the knots of additive and surface components jointly. This approach allows an unequal number of knots in the different covariates. In addition, separate shrinkage parameters for the linear, additive and surface parts of the model are allowed. Further, this approach permitted separate shrinkage parameters for the \(p\) responses within each of the three model parts. In contrast to other spline-based models, this approach allows the knot locations to move freely in the regressor space, and thus fewer knots are usually required. The estimation and computation details can be found in \textcite{li2013efficient}.

\section{Application to the M4 competition data}
\label{results}

\subsection{Dissimilarity between different datasets}
\label{dissimilarity-between-different-datasets}

Principal component analysis preserves the dissimilarity between widely separated data points rather than the similarity between nearby data points. This feature is useful for improved confidence in simulated series' representativeness of real time series. For example, if the simulated time series results in isolated clusters or highly dense clusters far apart from the real-world time series, it may indicate a poor representation of the real data.

We use principal component analysis to visualise dissimilarity between the datasets: observed time series (M1 and M3), simulated series and new time series (M4) in the two-dimensional instance space. The data are normalised using the Z-scale transformation before applying PCA.  The purpose of projecting M4 on the PCA-space created by the reference time series (M1, M3 and \textsf{GRATIS}) is to explore the differences in feature distribution characteristics between the training set (reference set) time series and the test set time series (M4). Since the meta-learner is trained on the reference data, the new time series to be forecast are required to lie within the space of the reference data. Therefore,  we first compute the principal component projection using the features in the reference set, and then project the new time series to the same low-dimensional feature space. The coverage of the reference set over the new time series in the space provides an initial idea about the suitability of the meta-learner in forecasting new time series. For example, if many series fall outside of the space covered by the reference set a new meta-learner is required to be trained.

\begin{figure}
  \centering
  \includegraphics[width=\textwidth]{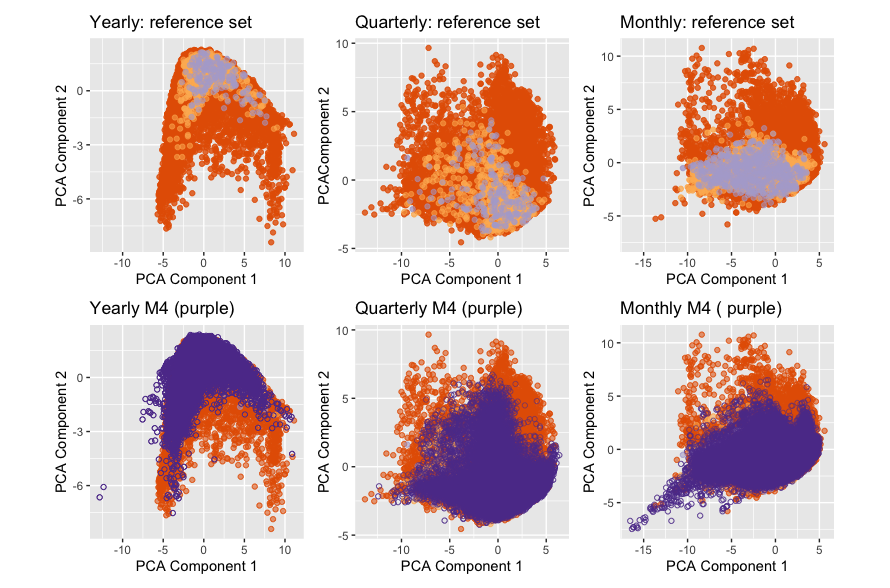}
  \includegraphics[width=\textwidth]{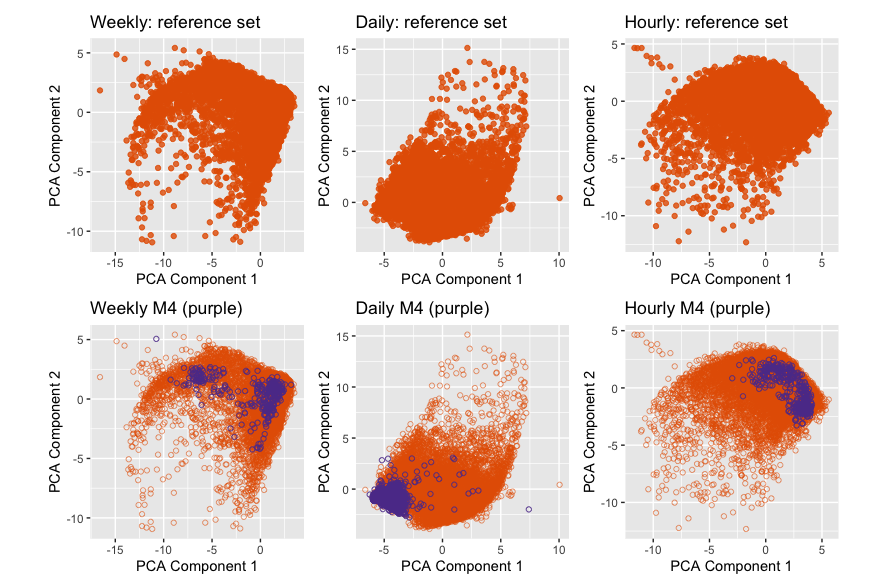}
  \caption{Distribution of yearly, quarterly, monthly, weekly, daily and hourly time series on the PCA space. On each subfigure, dark orange represents the simulated series and purple denotes new time series (M4 data). In the first row observed time series in the reference are highlighted in light orange and light purple. PCA space is computed based on the time series in the reference set and then the M4 competition (new series) are projected into the two-dimensional PCA space. Except for a few series, the majority of new time series we need to forecast fall within the space of the reference set. Note that for yearly, quarterly and monthly series the size of the test set is much larger than the size of the corresponding reference set.}
  \label{fig:pca}
\end{figure}

\begin{figure}
  \centering
  \includegraphics[width=\textwidth]{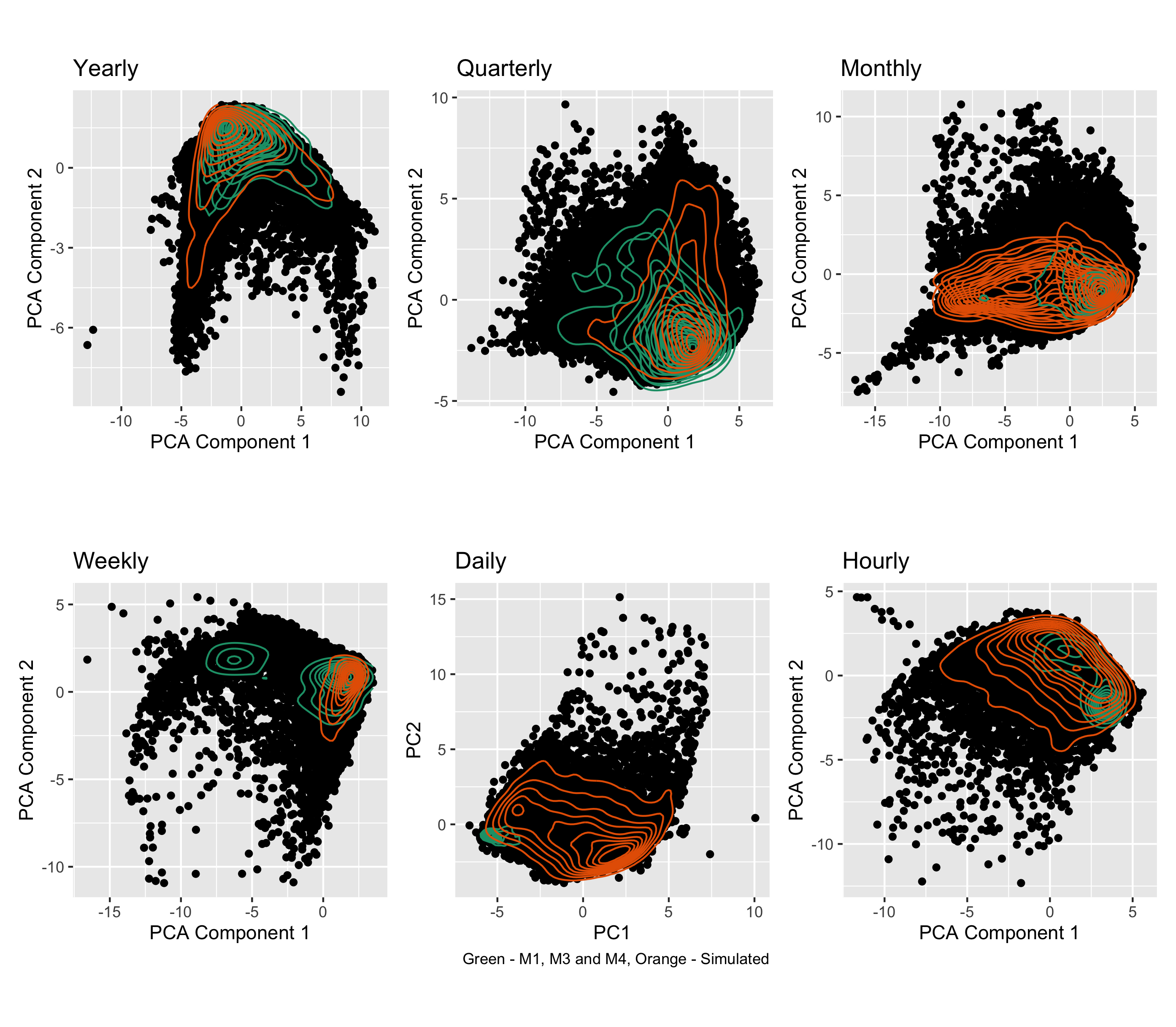}
  \caption{Visualisation of densities of yearly, quarterly, monthly, weekly, daily and hourly time series on the PCA space. On each graph, dark orange represents the simulated series and green denotes new time series (M4 data). Note that for yearly, quarterly and monthly series the size of the test set is much larger than the size of the corresponding reference set.}
  \label{fig:pcadensity}
\end{figure}

The projection also helps us to gain an idea about the global structure of the location of the different collections. The results are shown in \autoref{fig:pca}. The associated density plots are shown in \autoref{fig:pcadensity}. We compute principal components using the time series in the reference set (observed series and simulated series) and project new series (M4) into the two-dimensional space spanned by the first two eigenvectors. The first two principal components explain 51.3\%, 48.7\% and 48.3\% of the total variation in the yearly, quarterly and monthly data. We see that the distribution of the simulated time series (represented by the dark orange dots) clearly nests and fills in the instance space, although the simulated data are independent from the observed series except that we use the length distribution of the M4 series as a reference of that of the simulated series lengths, which is consistent with our theoretical expectation. Further, we can see that the projection of M4 series falls within the space created by the series in the reference set (M1, M3 and simulated). This is very important because our \textsf{FFORMPP} framework is trained based on the series in the reference set; hence, the model is valid over the space of the reference set. A few M4 time series in the monthly frequency category fall outside the space covered by the reference set in the first two principal components owing to very high lengths. Note that for yearly, quarterly and monthly series, the size of the M4 time series collection is much greater than the size of the corresponding collection of simulated series. For yearly and quarterly data, the number of time series in the M4 collection is about twice as large as the simulated series, and for monthly data, the M4 competition collection is four times larger than the simulated series. This shows the efficiency of the \textsf{GRATIS} simulation approach in increasing the diversity of feature space without having many time series similar in size to the new time series collection from which we wish to produce forecasts.

\begin{figure}
  \centering
  \includegraphics[width=\textwidth]{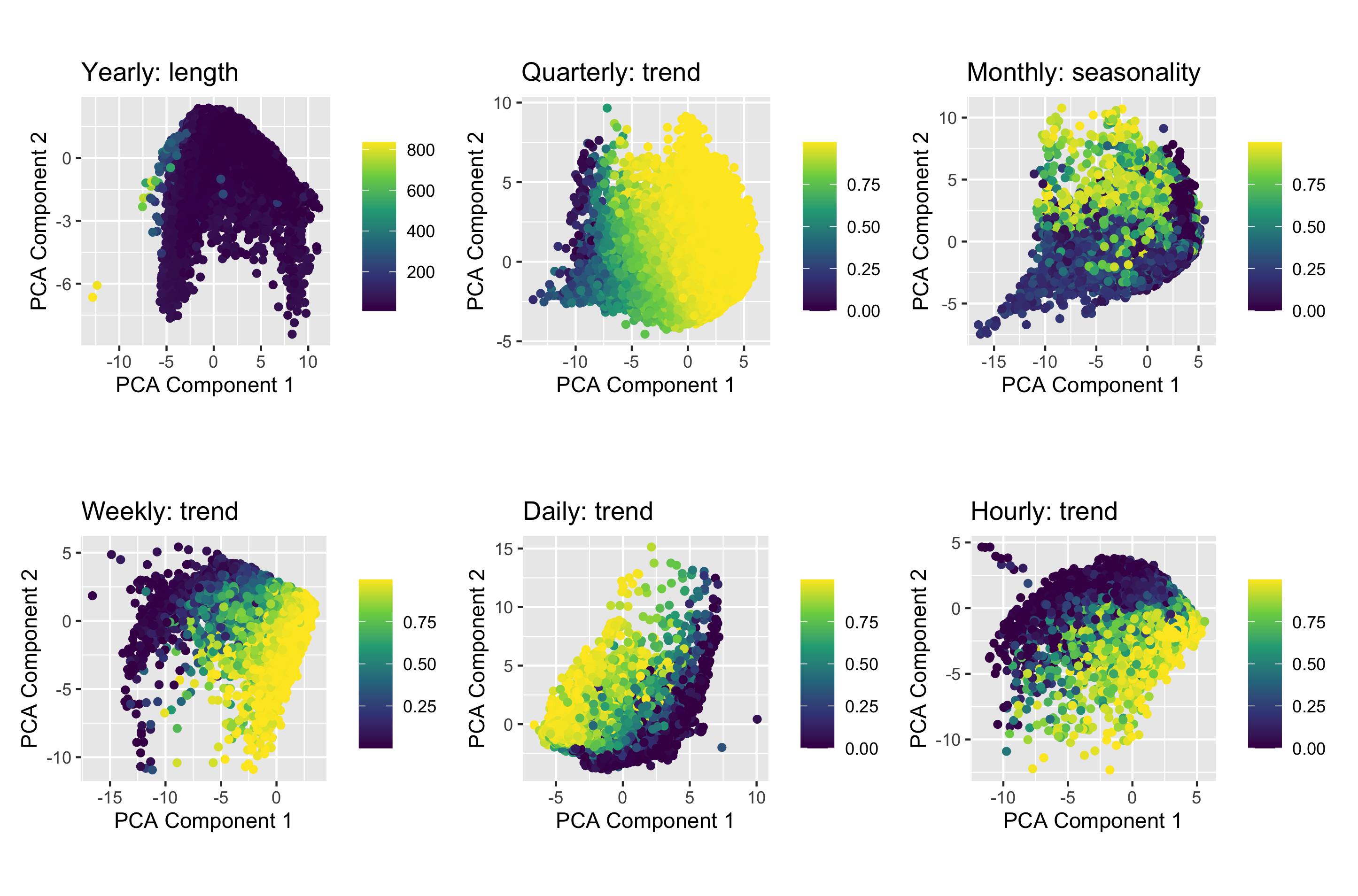}
  \caption{Visualisation of reasons that some test series of M4 fall outside the space of
the training set. For yearly series, the reason is high length values compared to training test, for the quarterly series high trend values, for monthly series low seasonality, and for weekly, daily and quarterly series the reason is heterogeneity in trend values. The legends on the right of the graphs show the values of the corresponding time series features.}
  \label{fig:reason}
\end{figure}

For weekly, daily and hourly time series the first two principal components explain 48.7\%, 43.4\% and 45.4\% of the total variance of weekly, daily and hourly series, respectively. In all three frequency categories, M4-competition data fall within the feature space of simulated series. For weekly, daily and hourly frequency categories, the simulated time series successfully fill in the space of the new time series. This shows the efficiency of the \textsf{GRATIS} approach in generating a realistic set of representative time series. However, the daily series of M4 competition series are clustered into a single location of the feature space. This is due to the quality issues of the M4 competition data. Daily series in the M4 competition appear very similar to each other owing to data leakages, as pointed out by~\citet{ingel2020correlated}. Examples of data leakage include the use of different segments of time series to create a new time series, and the addition of a constant value to create new time series.  Another reason for the series to be clustered in one place is the selected features are not diverse enough to capture the subtle changes of those time series.

\autoref{fig:reason} visualises the reasons that some test series of M4 fall outside the space of the training set. For year series the reason is high length values compared to the training set, for quarterly series high trend values, for monthly series low seasonality, and for weekly, daily and quarterly series the reason is heterogeneity in trend values. These associated features are identified by reviewing the coefficients of the PCA components. \autoref{fig:reason} is obtained by colouring the PCA feature space using the corresponding features. The information revealed from \autoref{fig:reason} can be taken into account in the offline phase of the algorithm to make the simulated
space even more representative of the M4 test set. For example, a few yearly M4 test series fall outside the training space, due to high length values. We can use the \textsf{GRATIS} approach to simulate time series that cover the full length range of the test data. Furthermore, for monthly series,  the values for the strength of seasonality of the training series are very high compared to actual ones. By generating series with strength of seasonality varying between 0.1 to 0.3, our training dataset will be more representative of the test set.

\subsection{Forecasting results}
\label{forecasting-results}

We evaluate the out-of-sample performances of our proposed meta-learning framework to the benchmarks based on the M4 competition series. MASE is used to evaluate forecast accuracy. We compute both individual and combination forecasts for the M4 competition series. Individual forecasts, denoted as ``FFORMPP-selection'', are computed based on the model corresponding to the minimum predicted MASE. The combination forecasts, denoted as ``FFORMPP-combination-4'', are computed by taking the median of individual forecasts corresponding to the four models with the minimum predicted MASE values. The reason for using the average of the best four models is, as pointed by \autocites{lichtendahl2013better}{petropoulos2020simple}, the simple average combination is as accurate as complex approaches and is a robust combination method that is hard to beat. Also, using the median of an even number of forecasts makes the final forecasts actually a combination of two individual forecasting methods, which simplifies the computation. The overall accuracy of each approach is computed using the weighted average based on the number of series involved in each data frequency. We also included the median forecast calculations based on 2, 4, 6, 8, and 10 models in order to show how forecasting accuracy affects when the number of models used increase. According to \autoref{forecast}, we can see forecasting accuracy based on the median of the 4 models with the minimum predicted MASE is higher than forecasting accuracy based on the median of the 2 models with the minimum MASE or the median of the 10 models with the minimum MASE. There is no significant difference between the overall forecast accuracy when we increase the number of models used from 4 to 6 or 8. Hence, our final combination forecasts are computed based on the median of the 4 models with the minimum MASE to reduce the computational cost.

We also compare our results with the individual forecasting models, their simple combinations including SA-median-12 (simple averaging using the median of all 12 models available) and SA-mean-12 (simple averaging using the mean of all 12 models), and \textsf{FFORMS}. \textsf{FFORMS} is trained based on the random forest algorithm. \textsf{FFORMS} outperforms \textsf{FFORMPP}-selection forecasts. However, \textsf{FFORMPP-combination-4} outperforms \textsf{FFORMS}. One could argue in the \textsf{FFORMS} algorithm, the random forest probability scores could be used for weighting alternative forecast models, constructing a robust combination scheme, or selecting a set of models for combination forecast. However, this approach did not bring any improvement in the performance. Furthermore, with some series, it degraded the performance. The reason is that in \textsf{FFORMS}, the forecast model selection problem is treated as a classification problem. Hence, in this case, the weights are mostly close to either 0 and 1; the best model has a weight close to 1 and others very close to 0. For example, suppose for a given series, the forecast error vector for the random walk, the random walk with drift and a white noise process is [1.31, 1.30, 3.4]. Then, ideally, the vector of class weight we expect from \textsf{FFORMS} is [0, 1, 0], i.e., the best model is given class probability 1 and others 0. This limitation is another motivation for us to introduce \textsf{FFORMPP}.

Further, we compare our results with those of the top three places of the M4 competition. The winning method of the competition is based on the Hybrid approach, which is a combination of exponential smoothing models and neural network approaches \autocite{makridakis2018m4}. The second and third approaches are based on combination forecasts computed based on nine and seven individual forecast models, respectively. According to the results of \autoref{forecast}, we can see that our approach achieved comparable results in a much more cost- and time-effective manner because our combination forecasts are calculated based on four individual models. The performance of the proposed method is comparable to that of the top-ranked forecasting methods of the competition.

In addition to that, we also compare our results with the most recent feature-based meta-learning algorithms introduced by \textcite{Li2020} and \textcite{wang2021uncertainty}. \textcite{Li2020} is a meta-learning algorithm developed based on imaging, and \textcite{wang2021uncertainty} is a feature-based algorithm developed based on Generalized Additive Models (GAMs). We can see from \autoref{forecast} that FFORMPP yields comparable accuracy results.

To conclude, the results in \autoref{forecast} suggest that the forecast combination obtained through the \textsf{FFORMPP} framework outperforms the corresponding benchmarks and other commonly used methods of forecasting. Note that in this study, all evaluations are made
on out-of-sample with a fixed forecast origin and forecast horizon.  The reason for evaluating performance through a fixed forecast origin is that most of the research done in the related area and results of the M4 forecasting competition are evaluated based on a fixed origin.  Hence, using a fixed forecast origin makes it easy to compare with published work. Furthermore, the out-of-sample accuracy of each forecasting method can also be evaluated using point forecasts generated using a rolling forecast origin in future research comparisons.

\begin{table}
  \caption{MASE values calculated over the M4 competition data. ``FFORMPP-selection'' results are computed based on the model corresponding to the minimum predicted MASE. ``FFORMPP-combination-4'' results are computed by taking the median of individual forecasts corresponding to the four models with  the minimum predicted MASE values. The median forecast calculations based on 2, 4, 6, 8, and 10 models are also included. ``SA-median-12'' means simple averaging using the median of all 12 models available, while ``SA-mean-12'' demonstrates simple averaging using the mean of all 12 models.}
  \centering
  \label{forecast}
    \resizebox{\textwidth}{!}{
  \begin{tabular}{lrrrrrrr}
    \toprule
& Yearly    & Quarterly & Monthly   & Weekly    & Daily     & Hourly    & Overall+ \\
 \midrule
 \bf{FFORMPP-selection}                                     & 3.37      & 1.17      & 1.05      & 2.53      & 4.26      & 1.06      & 1.75     \\
 \bf{FFORMPP-combination-4}                        & 3.07 & 1.13 & \bf{0.89} & 2.46 & 3.62      & 0.96 & 1.57     \\

  FFORMPP-combination-2                                     & 3.25      & 1.15      & 0.96      & \bf{2.38}      & 4.18      & 1.09      & 1.67     \\
  FFORMPP-combination-6                                     & 3.14      & \bf{1.12}      & 0.90      & 2.51      & 3.35      & 1.00      & 1.57     \\
FFORMPP-combination-8                                       & \bf{3.06}      & 1.16      & 0.92      & 2.54      & 3.30 & 1.11      & 1.57     \\
  FFORMPP-combination-10                                    & 3.28      & 1.22      & 0.94      & 2.57      & 3.52      & 1.33      & 1.66     \\
 \textsf{auto.arima}                                        & 3.40      & 1.17      & 0.93      & 2.55      & -         & -         & 1.59     \\
 \textsf{ets}                                               & 3.44      & 1.16      & 0.95      & -         & -         & -         & 1.60     \\
 \textsf{theta}                                             & 3.37      & 1.24      & 0.97      & 2.64      & 3.33      & 1.59      & 1.69     \\
 \textsf{rwd}                                               & 3.07 & 1.33      & 1.18      & 2.68      & \bf{3.25}      & 11.45     & 1.78     \\
 \textsf{rw}                                                & 3.97      & 1.48      & 1.21      & 2.78      & 3.27 & 11.60     & 2.04     \\
 \textsf{nn}                                                & 4.06      & 1.55      & 1.14      & 4.04      & 3.90      & 1.09      & 2.03     \\
 \textsf{stlar}                                             & -         & 2.02      & 1.33      & 3.15      & 4.49      & 1.49      & 1.72     \\
 \textsf{snaive}                                            & -         & 1.66      & 1.26      & 2.78      & 24.46     & 2.86      & 2.66     \\
 \textsf{tbats}                                             & -         & 1.19      & 1.05      & 2.49      & 3.27 & 1.30      & 1.22     \\
 \textsf{wn}                                                & 13.42     & 6.50      & 4.11      & 49.91     & 38.07     & 11.68     & 8.45     \\
 \textsf{mstlarima}                                         & -         & -         & -         & -         & 3.84      & 1.12      & 3.59     \\
 \textsf{mstlets}                                           & -         & -         & -         & -         & 3.73      & 1.23      & 3.50     \\
 SA-median-12                                               & 3.29      & 1.22      & 0.95      & 2.57      & 3.52      & 1.33      & 1.66     \\
 SA-mean-12                                                 & 4.09      & 1.58      & 1.16      & 6.96      & 7.94      & 3.93      & 2.25     \\
 \textsf{FFORMS} (random forest)                            & 3.17      & 1.20      & 0.98      & 2.31      & 3.57      & \bf{0.84}      & 1.65     \\
 \midrule \multicolumn{7}{c}{M4 competition top three places}                                                                                      \\
 \midrule M4-1st                                            & 2.98      & 1.12      & 0.88      & 2.36      & 3.45      & 0.89      & 1.53     \\
 M4-2nd                                                     & 3.06      & 1.11      & 0.89      & 2.11      & 3.34      & 0.81      & 1.54     \\
 M4-3rd                                                     & 3.13      & 1.12      & 0.91      & 2.16      & 2.64      & 0.87      & 1.54     \\
  \midrule \multicolumn{7}{c}{Meta-learning approaches}                                                                                        \\
 \midrule Image-based features \citep{Li2020}               & 3.13      & 1.12      & 0.90      & 2.26      & 3.46      & 0.84      & 1.57     \\
 FUMA - features and GAM models \citep{wang2021uncertainty} & 3.01      & 1.14      & 0.91      & NA        & NA        & NA        & 1.47     \\
 \bottomrule
  \end{tabular}}
   \begin{tablenotes}
     \item \footnotesize $^+$Weighted average based on the number of series for yearly, quarterly, monthly, weekly, daily and hourly.
  \end{tablenotes}
\end{table}

\section{Understanding the \textsf{FFORMPP} meta-learner}

\subsection{Performance impact factors}

To evaluate if \textsf{GRATIS} simulation approach leads to a better forecast, we also
trained a meta learning model using series originating only from the M1 and M3
competitions and compared results against the model training including the
\textsf{GRATIS}-based simulated time series. 
In the experiments,
we keep the training set as independent as possible of any knowledge of the time series
patterns of the test set, since the independence of the test set to the training set is of
fundamental importance to determine a model's out-of-sample predictive ability. The time
series generated from the \textsf{GRATIS} approach are independent of test set (M4 data)
except that the length distribution of the M4 series is used as a reference to that of the
simulated series lengths.

We run four experiments to evaluate which part (addition of simulated time series or the EBMSR approach) of the algorithm helps improve the forecasting performance. Their details are as follows.

\begin{itemize}
\item \textbf{Experiment 1.} We use only M1 and M3 competition data as our reference set. Then, similar to the approach used by \textcite{Petropoulos2014}, we use a multiple linear regression approach to model the relationship between forecast errors and features. Separate regression models are fitted to model the relationship between features and errors for different forecasting methods.

\item \textbf{Experiment 2}. In addition to M1 and M3 competition data, we also use a simulated time series. Multiple linear regression approach is used to model the relationship between and features. The only difference between the first and the second experiment is the addition of the simulated time series.

\item \textbf{Experiment 3}. We use M1 and M3 as our reference set. The EBMSR approach is used to model the forecast errors simultaneously as a function of features.

\item \textbf{Experiment 4}. It corresponds to our \textsf{FFORMPP} approach. We use M1, M3 and the simulated time series as our reference set. The EBMSR approach is used to model the forecast errors simultaneously as a function of features.
\end{itemize}

The accuracies of the four experiments are evaluated by comparing out-of-sample MASE. The
results of the experiments are shown in \autoref{rmse}. The overall accuracy is the
weighted average based on the number of series involved in yearly, quarterly and monthly
series. It can be seen that the accuracy in Experiment 2 is higher than that in Experiment
1. Furthermore, the accuracy of results in Experiment 4 (\textsf{FFORMPP}) is higher than
the accuracy of results in Experiment 3.  This suggests that the addition of the simulated
time series helps in improving the accuracy of the results. According to the results of
Experiment 1 and Experiment 3, using EBMSR over MLR when considering only the M1 and M3
data has a negative effect (reducing the forecasting accuracy). However, according to the
results of Experiment 4, combining EBMSR with the simulated data there is an improvement
in the forecasting accuracy. Hence, from the four experiments implemented, we find that
the effect of \textsf{GRATIS} depends on the algorithm used for employing the meta-learner
and using \textsf{GRATIS} and EBMSR separately does not always guarantee improvement
in forecasting accuracy. It could be the case that EBMSR requires a large collection of
series to identify the relationship between independent and dependent variables and that,
when this is not the case, simpler meta-learning algorithms may lead to better results.
The overall results inform that both simulated series and the EBMSR approach contribute to
the improvement of the forecasting accuracy.

\begin{table}
  \caption{MASE over the M4 Competition Data. The same experimental set-up introduced in
    Section \ref{results} is used for producing the results, with the only difference being the
    reference set considered for training the meta-learner algorithms.}
  \centering
  \label{rmse}
  \resizebox{\textwidth}{!}{
\begin{tabular}{llrrrrrrr}
  \toprule
                                             & Reference set       & Yearly & Quarterly & Monthly & Weekly & Daily & Hourly & Overall$^+$        \\
 \midrule                                    &                     & \multicolumn{6}{c}{Multiple Linear Regression (MLR)}                           \\
  \cline{3-9}
  Experiment 1                               & M1 \& M3            & 3.38   & 1.60      & 1.12    & -       & -      & -       & 1.75               \\
 Experiment 2                                & M1, M3 \& Simulated & 3.38   & 1.18      & 1.05    & 2.61    & 3.32   & 1.19    & 1.68               \\
 \cline{3-9}                                                                                                                                        \\
                                             &                     & \multicolumn{6}{c}{Efficient Bayesian Multivariate Surface Regression (EBMSR)} \\
  \cline{3-9} Experiment 3                   & M1 \& M3            & 3.44   & 1.18      & 1.47    & -       & -      & -       & 1.87               \\
 \cline{3-9} Experiment 4 (\textsf{FFORMPP}) & M1, M3 \& Simulated & 3.07   & 1.13      & 0.89    & 2.46    & 3.62   & 0.96    & 1.47               \\
  \bottomrule
\end{tabular}}
\begin{tablenotes}
\item \footnotesize $^+$Weighted average based on the number of series for yearly, quarterly and monthly.
\end{tablenotes}
\end{table}

To ensure a fair comparison, the computational time for producing forecasts based on 100 randomly selected series from each frequency category of the M4 competition data set is given in \autoref{forecasttime}. The reported values are the median elapsed time of 100 replicates. The corresponding Inter Quartile Ranges (IQRs) are given in parentheses. We can see that \textsf{FFORMPP} is significantly faster than the total computational time of the individual methods used for forecast combination. The computational time was measured using the \proglang{R} package \pkg{microbenchmark} \citep{microbenchmark} on 24 core Xeon-E5 2.50GHz servers.

\begin{table}[!h]
  \centering
\caption{Computational time for producing forecasts based on 100 randomly selected series from each frequency category of the M4 data set. The reported values are median elapsed time of 100 replicates. The corresponding Inter Quartile Ranges (IQRs) are given in parentheses.}
\label{forecasttime}
\resizebox{\textwidth}{!}{
\begin{tabular}{lrrrrrr}
    \toprule
    \multicolumn{7}{c}{Computational time for producing forecasts in seconds (IQR)}                                   \\
    \midrule
                        & Yearly      & Quarterly     & Monthly       & Weekly        & Daily         & Hourly        \\
    \midrule
    \textsf{FFORMPP}    & 5.31(0.21)  & 30.45 (1.44)  & 190.05 (5.15) & 183.36 (6.78) & 93.76 (0.34)  & 56.14 (11.67) \\
    \midrule
    \textsf{auto.arima} & 5.91 (0.05) & 42.11 (2.15)  & 448.41 (1.95) & 584.98 (2.35) & -             & -             \\
    \textsf{ets}        & 1.14 (0.02) & 16.92 (0.09)  & 115.50 (0.17) & -             & -             & -             \\
    \textsf{theta}      & 2.74 (2.48) & 10.15 (11.45) & 29.13 (1.15)  & 96.06 (0.42)  & 83.77 (2.67)  & 54.32 (2.32)  \\
 \textsf{rwd}           & 0.29 (5.42) & 0.29 (8.20)   & 0.33 (15.57)  & 0.34 (21.78)  & 0.41 (33.72)  & 0.37 (26.46)  \\
 \textsf{rw}            & 0.16 (4.65) & 0.20 (6.67)   & 0.26 (15.68)  & 0.22 (17.16)  & 0.27 (19.56)  & 0.24 (9.58)   \\
 \textsf{nn}            & 2.32 (0.14) & 6.54 (0.23)   & 32.78(0.25)   & 281.68 (0.61) & 424.28 (1.71) & 354.97 (3.61) \\
 \textsf{stlar}         & -           & 0.83 (17.97)  & 0.94 (12.03)  & 0.90 (10.11)  & 2.21 (0.12)   & 1.70 (0.01)   \\
 \textsf{snaive}        & -           & 0.18 (4.51)   & 0.30 (3.12)   & 0.20 (1.44)   & 0.32 (0.34)   & 0.44 (2.12)   \\
 \textsf{tbats}         & -           & 20.16 (6.98)  & 38.12 (2.44)  & 40.16 (3.36)  & 68.73 (0.65)  & 49.52 (2.98)  \\
 \textsf{wn}            & 0.19 (2.65) & 0.20 (4.30)   & 0.23 (0.08)   & 0.19 (4.51)   & 0.26 (1.00)   & 0. 22 (0.05)  \\
 \textsf{mstlarima}     & -           & -             & -             & -             & 86.92 (0.52)  & 30.60 (0.17)  \\
 \textsf{mstlets}       & -           & -             & -             & -             & 19.79 (0.09)  & 10.13 (0.48)  \\
\bottomrule
\end{tabular}
}
\end{table}

\subsection{Visualising patterns learned by the meta-learner}

\begin{figure}
\centering
\includegraphics[width=\textwidth]{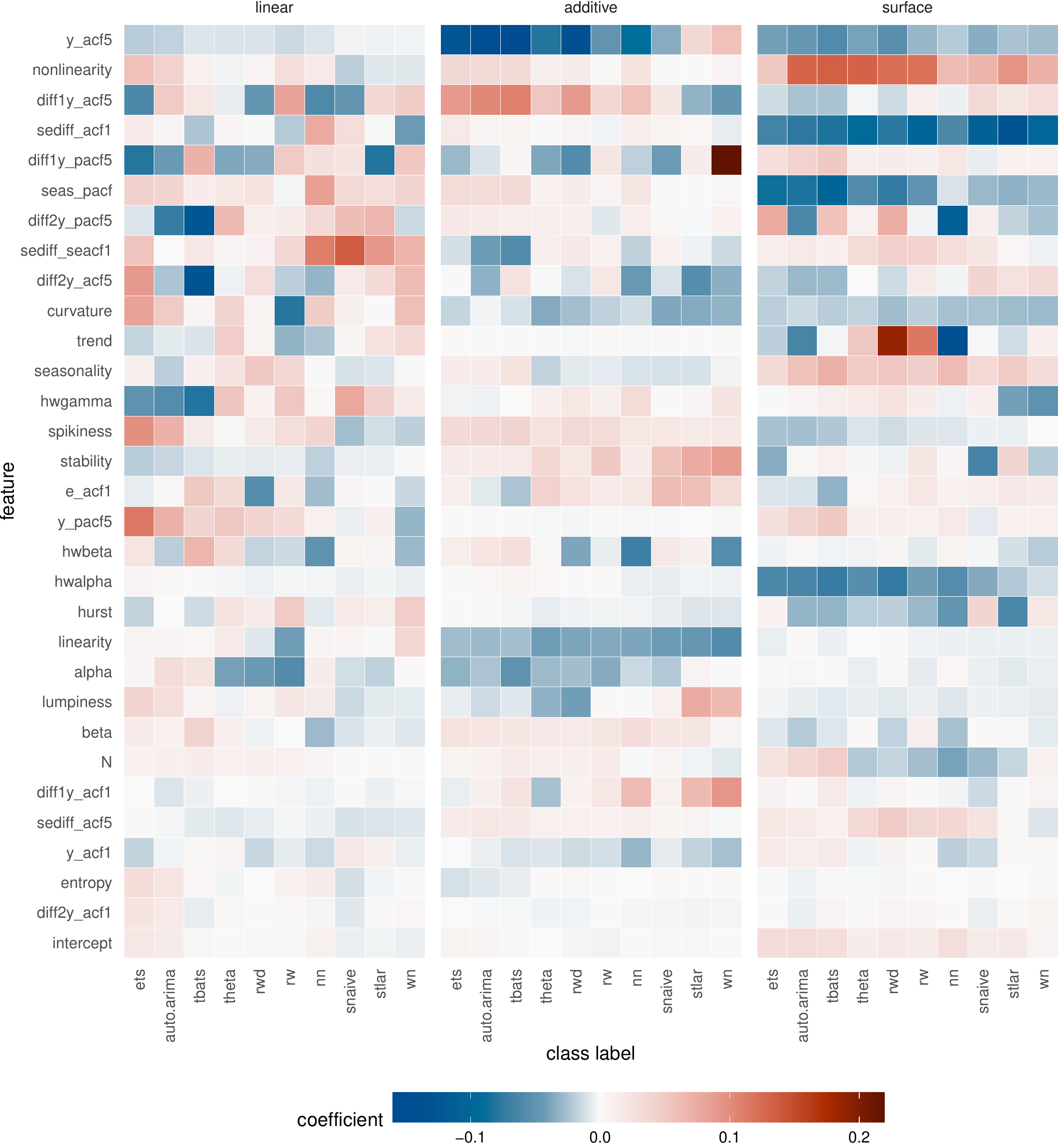}
  \caption{Visualisation of estimated model coefficients for the quarterly series. The Y-axis represents features, the X-axis represents class labels and the heat map cells represent estimated model coefficients. The class labels are ordered with their descending ranks.}
  \label{fig:qcoef}
\end{figure}

We first visualise the estimated model coefficients for the linear component, and two
nonlinear components (additive and surface) of the model. \autoref{fig:qcoef} shows the
heatmap of estimated model coefficients for the quarterly series. The corresponding
visualisations for yearly, monthly, weekly, daily and hourly are given in the
\autoref{appendix}.

According to \autoref{fig:qcoef}, the estimated coefficients of features related to
autocorrelation and partial autocorrelation coefficients drastically change across class
labels and within the three different components of the model. The estimated coefficients
related to trend, hurst exponent, nonlinearity, length and strength of seasonality vary
across classes labels within additive component of the model. Furthermore,
\autoref{fig:qcoef} shows that different features play a different role within the linear,
additive and surface components of the model. Zooming into individual cells, conditional
on other components and features, we can see within the additive component, as the feature
\texttt{y\_acf5} increases unit the estimated MASE value of \texttt{auto.arima} decreases,
while for \texttt{wn}, as \texttt{y\_acf5} increases the estimated MASE increases. That
indicates for series with high \texttt{y\_acf5} value, \texttt{auto.arima} algorithm gives
low MASE values, while the white noise process gives high MASE values. The other
coefficients can also be interpreted similarly. For general guidance of interpreting
semiparametric models as we used in the paper, we suggest to refer to
\citet{ruppert2003semiparametric} and \citet{li2013efficient}.

We also report the relative frequencies with which each forecast model is selected as a component of the calculation of combination forecasts based on \textsf{FFORMPP}. Further, to gain an idea of the different types of model combinations used to compute the forecast, we cluster time series based on the models that are used to calculate \textsf{FFORMPP} combination. For this purpose, we first create a design matrix of 1 and 0. The columns of the design matrix correspond to each forecast model, and rows correspond to each time series. The cell values of the matrix are 1s and 0s, where 1 is assigned if a corresponding model is used for combination forecast, and 0 otherwise. Then, hierarchical clustering was done by using a binary distance metric and ward clustering method. A cluster analysis was performed separately for each frequency category.

\begin{figure}
  \centering
  \includegraphics{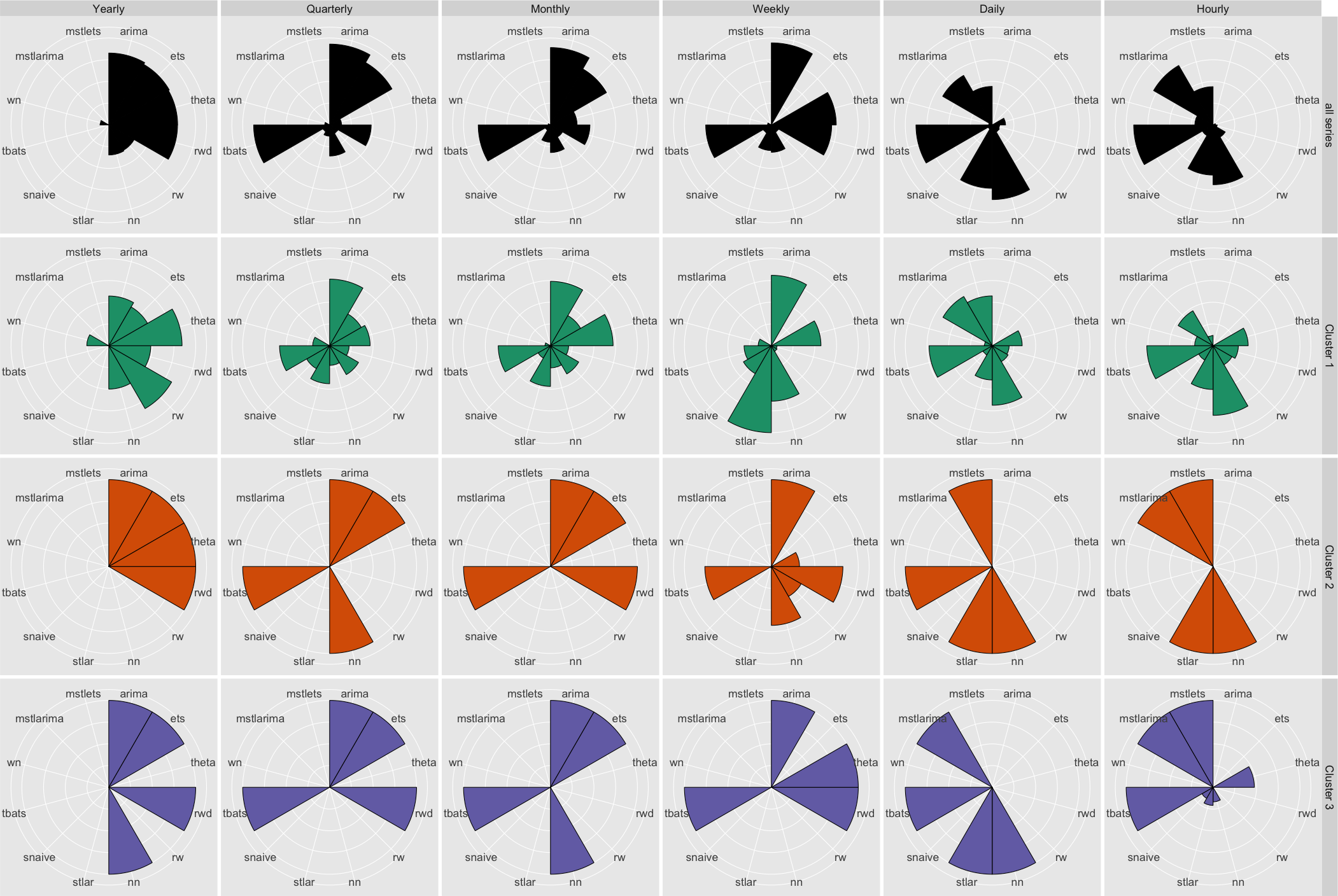}
  \caption{\label{fig:radarplot}Visual representation of relative frequencies that each forecast model was selected as a component to the calculation of combination forecasts based on \textsf{FFORMPP}. Each polar coordinate shows relative frequency that each forecast model was selected as a component to the calculation of combination forecasts.}
\end{figure}

For each frequency category, we identified three main clusters. The results are shown in \autoref{cluster}, and the associated graphical representation is shown in \autoref{fig:radarplot}. According to the results of \autoref{forecast}, for yearly data we can see that \texttt{auto.arima}, \texttt{ets}, \texttt{theta} and \texttt{rwd} give the best individual forecast. From \autoref{cluster}, we can see that those four models were most frequently selected to the \textsf{FFORMPP} combination forecast. Similarly, from \autoref{forecast} we can see that for the quarterly and monthly series \texttt{auto.arima}, \texttt{ets} and \texttt{tbats} provide the best individual forecasts, and according to \autoref{cluster}, we can see that those models are selected most often (approximately greater than 75\%) for the quarterly and monthly series. Similarly, we can interpret the results for weekly, daily and hourly series. Cluster 3 in the yearly series is very similar to cluster 2, the only difference being that, cluster 3 uses \texttt{nn} instead of \texttt{theta} model. The series in the first cluster uses a different combination of models from that which we considered for the yearly series, apart from the two combinations used in cluster 2 and cluster 3, respectively. for the quarterly series, the biggest cluster is cluster 3, in which \texttt{auto.arima}, \texttt{ets}, \texttt{rwd} and \texttt{tbats} are used to calculate combination forecasts. Similar to the results of quarterly series, for monthly and daily series we observe two clusters that are homogeneous in terms of models used to compute combination forecasts. In terms of weekly data, most of the series use \texttt{auto.arima}, \texttt{theta}, \texttt{rwd} and \texttt{tbats} for combination forecasts. For daily and hourly series, it is interesting to observe that all the series in cluster 2 and cluster 3 use at least one of the models \texttt{mstlarima} or \texttt{mstlets}, which handle multiple seasonality, as a component in calculating combination forecast.

\begin{table}
  \caption{Relative frequencies that each forecast model was selected as a component to the calculation of combination forecasts based on \textsf{FFORMPP} (All values are shown in percentages).}
  \label{cluster} \resizebox{\textwidth}{!}{
    \begin{tabular}{lrrrrrrrrrrrrr}
      \toprule
      Source        & {No. of series} & \textsf{auto.arima} & \textsf{ets}  & \textsf{theta} & \textsf{rwd}  & \textsf{rw} & \textsf{nn}   & \textsf{stlar} & \textsf{snaive} & \textsf{tbats} & \textsf{wn} & \textsf{mstlarima} & \textsf{mstlets} \\
 \midrule Yearly    & 23000           & \textbf{82.7}       & \textbf{80.4} & \textbf{79.2}  & \textbf{79.1} & 33.7        & 34.6          & -              & -               & -              & 10.2        & -                  & -                \\
 ~~~~Cluster 1      & 9287            & 57.2                & 51.5          & 84.3           & 48.3          & 83.5        & 49.9          & -              & -               & -              & 25.3        & -                  & -                \\
 ~~~~Cluster 2      & 10391           & 100                 & 100           & 100            & 100           & 0           & 0             & -              & -               & -              & 0           & -                  & -                \\
 ~~~~Cluster 3      & 3322            & 100                 & 100           & 0              & 100           & 0           & 100           & -              & -               & -              & 0           & -                  & -                \\
 \midrule Quarterly & 24000           & \textbf{93.1}       & \textbf{82.8} & 13.8           & 47.8          & 11.4        & 35.8          & 12.9           & 9.1             & \textbf{87.4}  & 5.8         & -                  & -                \\
 ~~~~Cluster 1      & 7126            & 76.6                & 42.1          & 46.6           & 22.5          & 38.4        & 22.6          & 43.7           & 30.5            & 57.6           & 19.5        & -                  & -                \\
 ~~~~Cluster 2      & 6994            & 100                 & 100           & 0              & 0             & 0           & 100           & 0              & 0               & 100            & 0           & -                  & -                \\
 ~~~~Cluster 3      & 9880            & 100                 & 100           & 0              & 100           & 0           & 0             & 0              & 0               & 100            & 0           & -                  & -                \\
 \midrule Monthly*  & 48000           & \textbf{88.9}       & \textbf{74.4} & 30.6           & 45.4          & 16.1        & 31.8          & 19.9           & 7.1             & \textbf{83.1}  & 2.8         & -                  & -                \\
 ~~~~Cluster 1      & 12615           & 74.1                & 40.0          & 72.0           & 20.9          & 37.5        & 25.5          & 46.8           & 16.5            & 60.1           & 6.5         & -                  & -                \\
 ~~~~Cluster 2      & 11084           & 100                 & 100           & 0              & 100           & 0           & 0             & 0              & 0               & 100            & 0           & -                  & -                \\
 ~~~~Cluster 3      & 6301            & 100                 & 100           & 0              & 0             & 0           & 100           & 0              & 0               & 100            & 0           & -                  & -                \\
 \midrule Weekly    & 359             & \textbf{94.4}       & -             & \textbf{74.7}  & 69.4          & 9.8         & 31.5          & 29.5           & 10.9            & \textbf{75.5}  & 4.5         & -                  & -                \\
 ~~~~Cluster 1      & 105             & 80.9                & -             & 57.1           & 6.7           & 76.2        & 63.8          & 100            & 37.1            & 31.4           & 15.2        & -                  & -                \\
 ~~~~Cluster 2      & 68              & 100                 & -             & 32.4           & 82.4          & 39.7        & 67.6          & 1.5            & 0               & 76.5           & 0           & -                  & -                \\
 ~~~~Cluster 3      & 186             & 100                 & -             & 100            & 100           & 0           & 0             & 0              & 0               & 100            & 0           & -                  & -                \\
 \midrule Daily     & 4227            & -                   & -             & 18.4           & 8.7           & 9.5         & \textbf{85.9} & \textbf{72.9}  & 5.63            & \textbf{87.7}  & 4.1         & 65.8               & 44.3             \\
 ~~~~Cluster 1      & 1891            & -                   & -             & 34.5           & 19.5          & 21.2        & 68.5          & 39.9           & 12.6            & 72.6           & 9.1         & 65.4               & 52.3             \\
 ~~~~Cluster 2      & 791             & -                   & -             & 0              & 0             & 0           & 100           & 100            & 0               & 100            & 0           & 0                  & 100              \\
 ~~~~Cluster 3      & 1545            & -                   & -             & 0              & 0             & 0           & 100           & 100            & 0               & 100            & 0           & 100                & 0                \\
 \midrule Hourly    & 414             & -                   & -             & 2.9            & 4.8           & 16.2        & 68.8          & 57.7           & 14.3            & \textbf{91.3}  & 21.0        & \textbf{78.9}      & 43.9             \\
 ~~~~Cluster 1      & 151             & -                   & -             & 40.4           & 29.1          & 25.2        & 80.1          & 50.3           & 19.2            & 76.2           & 21.2        & 46.4               & 11.9             \\
 ~~~~Cluster 2      & 162             & -                   & -             & 0              & 0             & 0           & 100           & 100            & 0               & 0              & 0           & 100                & 100              \\
 ~~~~Cluster 3      & 101             & -                   & -             & 47.5           & 0             & 0           & 16.8          & 20.8           & 14.9            & 100            & 0           & 100                & 100              \\
 \bottomrule
    \end{tabular} }
    \begin{tablenotes}
    \item Note: *Cluster analysis is based on 30000 series. 
    \end{tablenotes}
\end{table}

\begin{figure} \centering \includegraphics{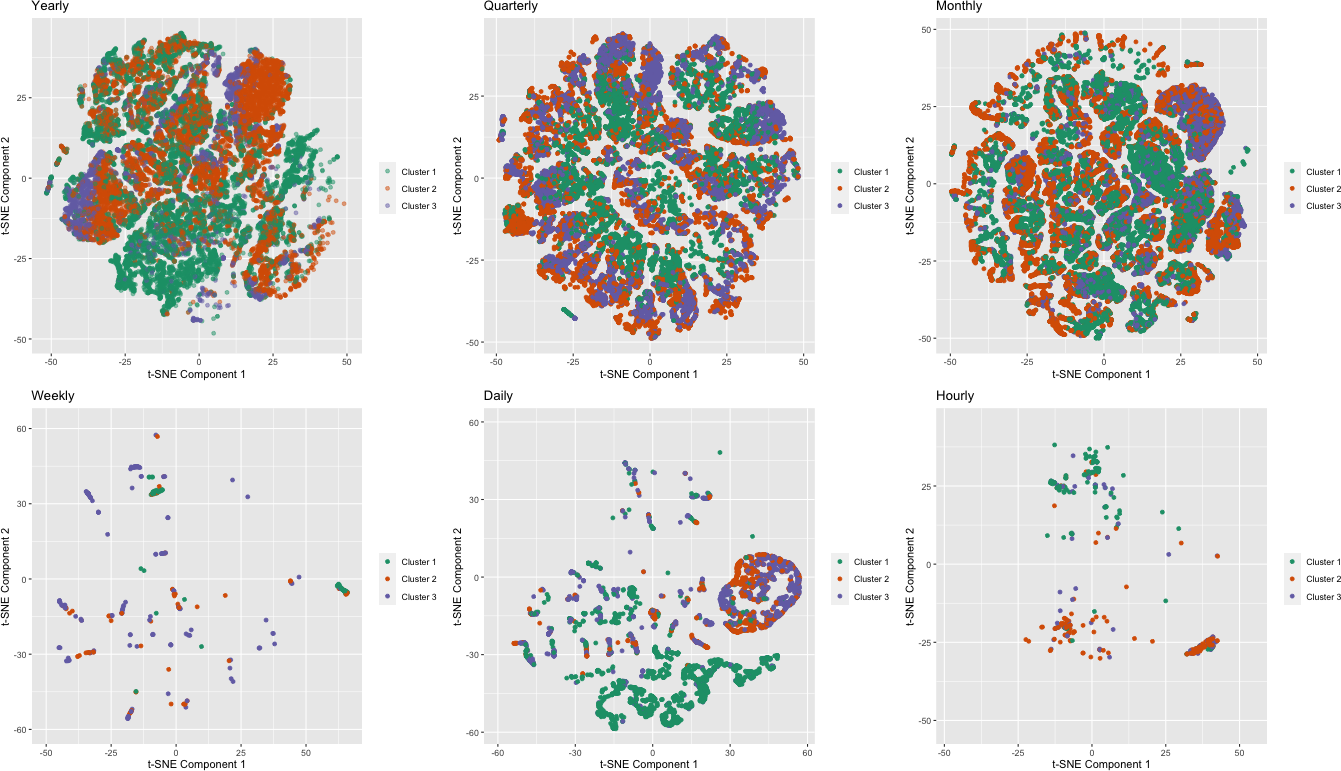}
  \caption{Location of the three clusters in the instance space. }
  \label{fig:clusterplot}
\end{figure}

We now examine the locations of different clusters of M4 competition series in the instance space defined by features. \autoref{fig:clusterplot} shows the locations of the three clusters in the instance space computed based on the t-SNE approach~\citep{Maaten2008}. For yearly series, we can see that most of the series in cluster 1 fall just below the diagonal, while series in clusters 2 and 3 fall within the upper triangular region. Further, clusters 2 and 3 are similar and the series corresponding to cluster 2 and 3 stay close together. for the quarterly series, an interesting pattern of clusters can be observed. The time series in cluster 1 fall within the inner circle, while the series in clusters 2 and 3 fall within the second and third outer rings of the circle. This pattern of clusters is called non-spherical. For monthly series, the grouping of the clusters is not obvious. However, a close inspection of the instance space shows a spiral pattern of clusters. Further, cluster 3 is preferred by the series in the upper right corner. For weekly series, clusters are more dispersed across the instance space. For daily and hourly series, clear separation of cluster 1 from clusters 2 and 3 can be observed. Clearly, within each frequency category, similar clusters (for example, clusters 2 and 3 for yearly, quarterly and monthly series) are located close together in the instance space, ensuring the similarity of the features of those time series.

\begin{figure} \centering \includegraphics[width=1.01\textwidth]{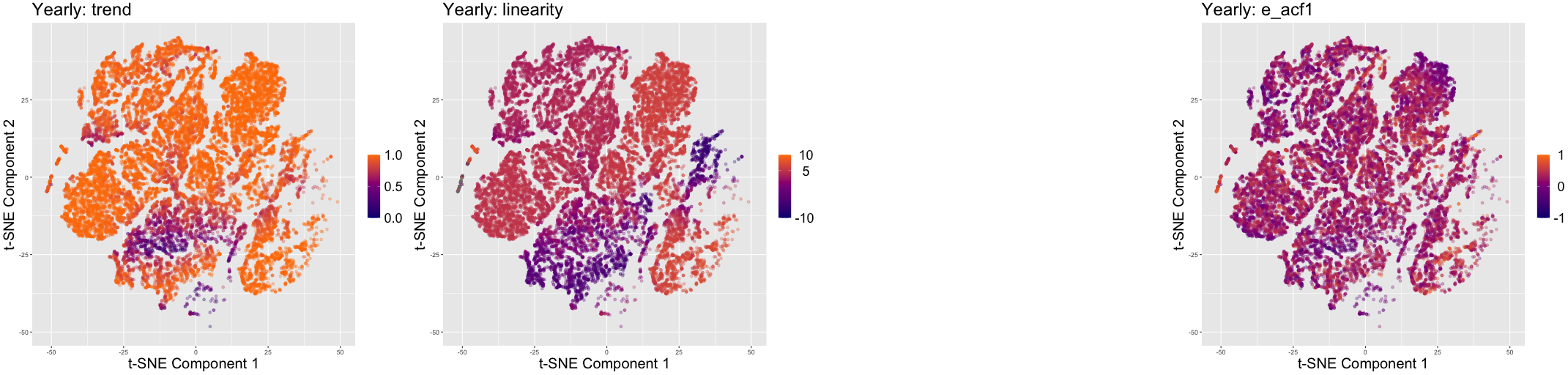}\\
 \includegraphics[width=\textwidth]{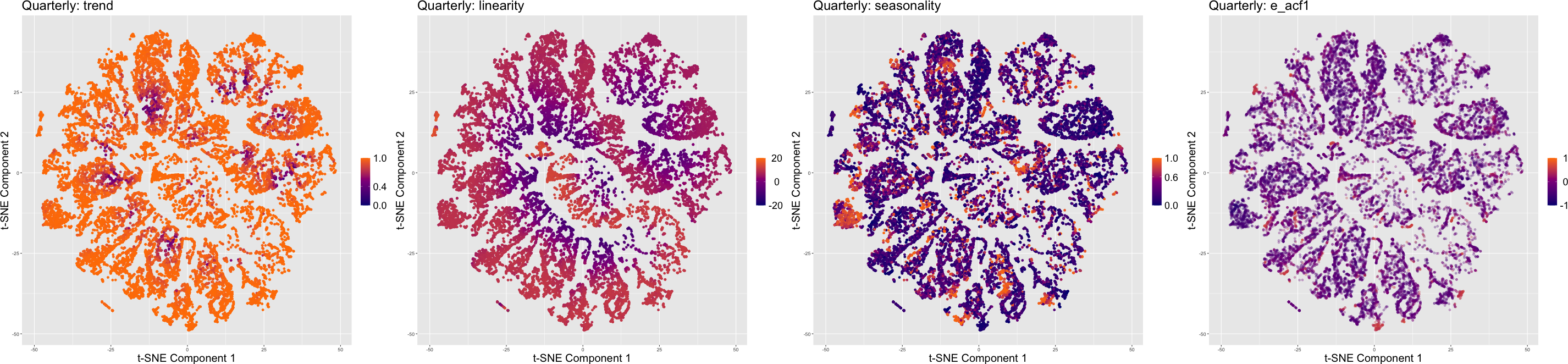}\\
 \includegraphics[width=\textwidth]{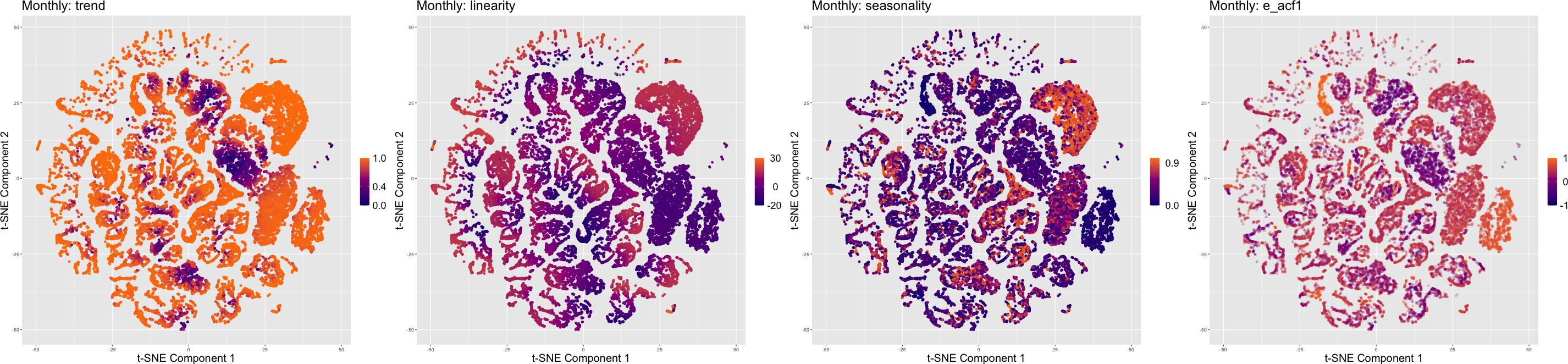}\\
 \includegraphics[width=\textwidth]{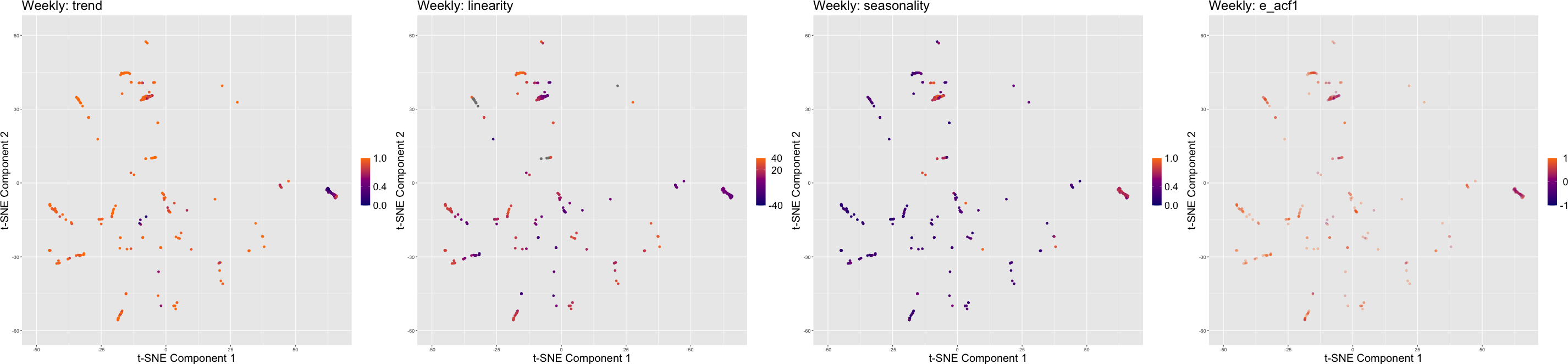}\\
 \includegraphics[width=\textwidth]{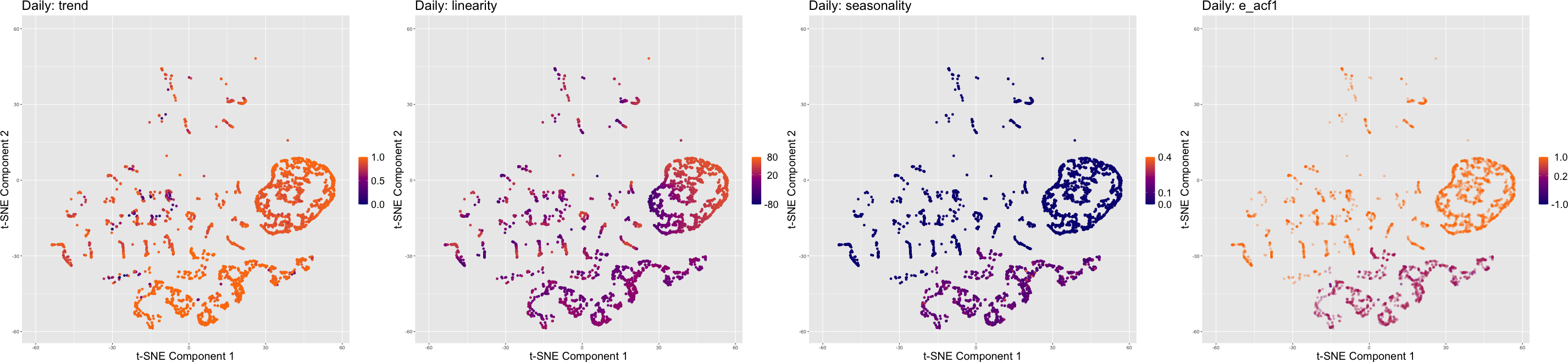}\\
 \includegraphics[width=\textwidth]{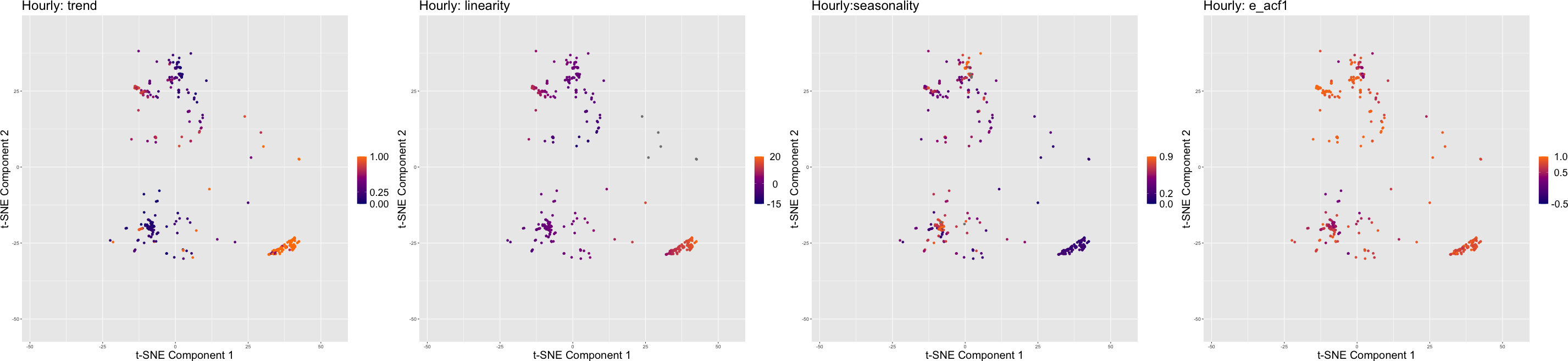}\\
  \caption{Distribution of features \texttt{trend}, \texttt{linearity} \texttt{seasonality} and \texttt{e\_acf1} across instance space of M4 competition data. The instances in the lower right corner of the instance space are highly trended and less seasonal. Note that the \texttt{seasonality} feature is not available for the yearly data.}
  \label{fig:featureplotshourly}
\end{figure}

The useful description provided by \autoref{fig:clusterplot} further prompts us to consider the challenge of identifying how the features of time series influence the grouping of these clusters. We now consider how different features vary across the instance space to understand how the locations of different time series reveal the relationship between the features and cluster separation. The preliminary results of the M4 competition \autocite{makridakis2018m4} show that randomness of time series is the most critical factor influencing the forecast accuracy, followed by linearity. Further, their follow-up paper \textcite{spiliotis2019forecasting} points out that highly trended and seasonal time series tend to be easier to forecast. The information about teh remainder series is useful for gaining an idea of the random variation not explained by the trend and seasonality of the series. Hence, we explore the instance space corresponding to the features, strength of trend (\texttt{trend}), the extend of linearity (\texttt{linearity}), strength of seasonality (\texttt{seasonality}) and the first autocorrelation coefficient of the remainder series after applying STL decomposition on the time series (\texttt{e\_acf1}) \autocite{cleveland1990stl}. The results are shown in \autoref{fig:featureplotshourly} for yearly, quarterly, monthly, weekly, daily and hourly series, respectively.

Combining the views of the instance spaces in \autoref{fig:clusterplot} and \autoref{fig:featureplotshourly}, we can have a picture of how the features contribute to the differences in clusters. According to \autoref{fig:featureplotshourly}, for yearly, quarterly, monthly and weekly series, highly trended series are more likely to fall within clusters 2 and 3. For yearly, quarterly and monthly series, most of the time series in the green cluster (in \autoref{fig:clusterplot}) are less trended with low \texttt{linearity} (in \autoref{fig:featureplotshourly}). Hence, those time series can be considered hard or challenging series to forecast. This is further confirmed from the results of \autoref{forecast} and \autoref{fig:clusterplot} -- cluster 1 is more heterogeneous according to the way the models are selected to compute combination forecasts (because they use different combinations of individual forecast models), in contrast to clusters 2 and 3. Furthermore, according to  \autoref{fig:featureplotshourly}, we can see that for yearly, quarterly and monthly series, the instance space coloured by linearity exhibits a similar structure to the corresponding cluster distribution across instance space shown in \autoref{fig:clusterplot}. For daily series, the features \texttt{seasonality} and \texttt{e\_acf1} clearly separate cluster 1 from the rest.  \autoref{fig:featureplotshourly} shows that, for hourly time series, features appear to separate the instance space into left and right.

\section{Conclusions}\label{conclusion}

This paper proposes a new meta-learning framework for large-scale time series
forecasting. The proposed framework can be used to compute both individual forecasts and
combination forecasts, which can also be easily generalised to allow for forecast selection, pooling or combination depending on the requirements and objectives of the forecasters. Moreover, the results provide useful insights regarding which forecasting models are expected to better for particular types of series. The proposed algorithm provides comparable accuracy results to that of top-ranked forecasting methods of the competition at lower computational cost, and can be easily parallelisable for use with large data sets for a given computing budget, which is important for supporting decisions. Furthermore, we present a visual guide to exploring the use of meta-learning for forecast model selection/combination and how the meta-learners are affected by various factors, such as the size of the selected method pool, and the size and diversity of the training samples.

Our results show that features of time series are useful in
selecting an optimal subset of models from all individual models without the need to run
all possible combinations of individual models during the online phase. Apart from the
obvious utility of this approach for forecast model selection, the ranking of models
provides an alternative solution to practitioners who may wish to incorporate their own
judgements or expertise into the forecasting decision process. We further explored the instance space defined by features to understand how certain
features of the time series are influencing the forecast model selection. A further
contribution of the paper is the provision of empirical support for the findings of the M4
competition \autocite{makridakis2018m4} that hold that the combination forecasts, in
general, outperform the best individual forecasts. An interesting future extension of this
framework would be to apply this methodology to produce probabilistic forecasts.

In general, the performance of any model strongly depends on the dataset (reference set)
used to train the model. We investigated the feasibility of using the \textsf{GRATIS} approach to
increase the diversity of the reference set. This approach is beneficial when researchers
have a small sample with which to build a reliable classifier, or no sample is available
due to data privacy issues. However, one could argue there can be parts of the simulation that are not very
representative of reality, and learning models might pay a cost for fitting this
unnecessary part of its model. The \textsf{GRATIS} approach simulates a realistic set of time
series by taking into account the natural boundaries of the features due to constraints on
combinations of features.  For example, it is not possible to have a time series with a very
low spectral entropy and a very low trend. By using \textsf{GRATIS}, we eliminate the possibilities
of generating time series that are not representative of reality.

The accuracy of \textsf{FFORMPP} individual model selection results are less accurate in yearly and
monthly time series than naively picking a single method that is known to work well a
priori such as Random Walk with Drift for yearly time series or auto.arima/ets/theta for
monthly. The possible reasons could be: i) the lack of diversity in the M4 competition
yearly, monthly time series, ii) features may not be discriminative enough to reveal the
uniqueness of the time series models, and iii) the efficient Bayesian multivariate surface
regression model is not suitable in identifying the best forecast model. This highlights
the need for the following further research.

\begin{itemize}
\item The current applications are limited to the M1, M3 and M4 competition
  datasets. Therefore, the applicability of the features space and algorithm space
  introduced in this paper are limited to the collection of time series of similar
  features as the data in the M-competitions. An important research direction is to expand
  the frameworks to other datasets that come from different application domains such as
  forecasting stock return data, electricity demand data, or irregular time series, etc.
  When adapting the frameworks to other applications the feature space should be revised
  with appropriate features that measure the characteristics of interest. The algorithm
  space may also be revised with suitable forecast models.
  Strictly speaking, any proposed approach after the availability of post-sample
  data, which may allow for testing and hyper-parametrisation, is not directly comparable
  with the methods submitted in the M4 competition, so does our method.

\item A vital component of a meta-learning framework is the construction of an engine that
  maps an input space composed of features to an output space consisting of forecast model
  performance. In fact, Bayesian spline regression with shrinkage estimator or spike and
  slab prior (standard setting) can be considered as an analogy to LASSO in regression
  problems. 
  A future research direction to investigate could be
  to replace the training algorithm with other alternatives such as deep-learning
  architectures, machine learning algorithms (e.g., random forest, XGBoost), etc., and
  test whether these approaches outperform the results given by Bayesian multivariate
  surface regression.  The \textsf{FFORMPP} framework is implemented in an R package
  \texttt{fformpp}, which can be downloaded from
  \url{https://github.com/thiyangt/fformpp}.

\end{itemize}

\section*{Acknowledgements}
\label{acknowledgements}

The authors are grateful to the editors, two anonymous
reviewers, Prof. Rob J Hyndman and Prof. George Athanasopoulos from Monash University for
helpful comments that improved the contents of the paper.

Thiyanga S. Talagala's research was supported by Monash University, Australia, the
Australian Research Council (ARC) the Centre of Excellence for Mathematical and
Statistical Frontiers (ACEMS). Yanfei Kang is supported by the National Natural Science
Foundation of China (No. 72021001) and the National Key Research and Development Program
(No. 2019YFB1404600). Feng Li is supported by the Beijing Universities Advanced
Disciplines Initiative (No. GJJ2019163) and the disciplinary funding of Central University
of Finance and Economics. This research was supported by Alibaba Group through Alibaba
Innovative Research Program and in part by the Monash eResearch Centre and
eSolutions-Research Support Services through the use of the MonARCH High Power Computing
(HPC) Cluster.

\printbibliography

@Book{ruppert2003semiparametric,
  author    = {Ruppert, D. and Wand, M.P. and Carroll, R.J.},
  publisher = {Cambridge University Press, Cambridge},
  title     = {{Semiparametric regression}},
  year      = {2003},
  isbn      = {0521785162},
}

@article{hall2009weka,
  title={The WEKA data mining software: an update},
  author={Hall, Mark and Frank, Eibe and Holmes, Geoffrey and Pfahringer, Bernhard and Reutemann, Peter and Witten, Ian H},
  journal={ACM SIGKDD explorations newsletter},
  volume={11},
  number={1},
  pages={10--18},
  year={2009},
  publisher={ACM New York, NY, USA}
}

@article{kang2020gratis,
  title={GRATIS: GeneRAting TIme Series with diverse and controllable characteristics},
  author={Kang, Yanfei and Hyndman, Rob J and Li, Feng},
  journal={Statistical Analysis and Data Mining},
  volume={13},
  number={4},
  year={2020}
}

@article{li2013efficient,
  title={Efficient Bayesian multivariate surface regression},
  author={Li, Feng and Villani, Mattias},
  journal={Scandinavian Journal of Statistics},
  volume={40},
  number={4},
  pages={706--723},
  year={2013},
  publisher={Wiley Online Library}
}

@Manual{forecast,
    title = {{forecast}: Forecasting functions for time series and linear models} ,author = {Rob Hyndman and George Athanasopoulos and Christoph Bergmeir and Gabriel Caceres and Leanne Chhay and Mitchell O'Hara-Wild and Fotios Petropoulos and Slava Razbash and Earo Wang and Farah Yasmeen},
year = {2018},
note = {R package version 8.3},
url = {http://pkg.robjhyndman.com/forecast},
}

@misc{tashman1991automatic,
  title={Automatic forecasting software: A survey and evaluation},
  author={Tashman, Leonard J and Leach, Michael L},
  journal={International Journal of Forecasting},
  volume={7},
  number={2},
  pages={209--230},
  year={2013},
  year={1991},
  publisher={Elsevier}
}

@article{collopy1992rule,
  title={Rule-based forecasting: development and validation of an expert systems approach to combining time series extrapolations},
  author={Collopy, Fred and Armstrong, J Scott },
  journal={Management Science},
  volume={38},
  number={10},
  pages={1394--1414},
  year={1992},
}

@article{adya2001automatic,
  title={Automatic identification of time series features for rule-based forecasting},
  author={Adya, Monica and Collopy, Fred and Armstrong, J Scott and Kennedy, Miles},
  journal={International Journal of Forecasting},
  volume={17},
  number={2},
  pages={143--157},
  year={2001},
}

@article{wang2009rule,
  title={Rule induction for forecasting method selection: meta-learning the characteristics of univariate time series},
  author={Wang, Xiaozhe and Smith-Miles, Kate and Hyndman, Rob J},
  journal={Neurocomputing},
  volume={72},
  number={10},
  pages={2581--2594},
  year={2009},
}

@article{shah1997model,
  title={Model selection in univariate time series forecasting using discriminant analysis},
  author={Shah, Chandra},
  journal={International Journal of Forecasting},
  volume={13},
  number={4},
  pages={489--500},
  year={1997},
  publisher={Elsevier}
}

@techreport{fforms,
  title={Meta-learning how to forecast time series},
  author={Talagala, Thiyanga S and Hyndman, Rob J and Athanasopoulos, George},
  type = {Working Paper},
  number = {6/18},
  institution = {Department of Econometrics \& Business Statistics, Monash University},
  year = {2018},
}

@article{rice1976,
  title={The algorithm selection problem},
  author={Rice, J R},
  journal={Advances in Computers},
  volume={15},
  pages={65-118},
  year={1976},
}

@inproceedings{prudencio2004using,
  title={Using machine learning techniques to combine forecasting methods},
  author={Prud{\^e}ncio, Ricardo and Ludermir, Teresa},
  booktitle={Australasian Joint Conference on Artificial Intelligence},
  pages={1122--1127},
  year={2004},
  organization={Springer}
}

@article{meade2000evidence,
  title={Evidence for the selection of forecasting methods},
  author={Meade, Nigel},
  journal={Journal of Forecasting},
  volume={19},
  number={6},
  pages={515--535},
  year={2000},
  publisher={Wiley Online Library}
}

@Article{Maaten2008,
  author  = {Maaten, Laurens van der and Hinton, Geoffrey},
  journal = {Journal of Machine Learning Research},
  title   = {Visualizing data using t-{SNE}},
  year    = {2008},
  number  = {Nov},
  pages   = {2579--2605},
  volume  = {9},
}

@Article{fforma,
  author  = {Pablo Montero-Manso and George Athanasopoulos and Rob J. Hyndman and Thiyanga S. Talagala},
  journal = {International Journal of Forecasting},
  title   = {{FFORMA}: Feature-based forecast model averaging},
  year    = {2020},
  issn    = {0169-2070},
  number  = {1},
  pages   = {86 - 92},
  volume  = {36},
}

@Manual{microbenchmark,
    title = {microbenchmark: Accurate Timing Functions},
    author = {Olaf Mersmann},
    year = {2019},
    note = {R package version 1.4-7},
    url = {https://CRAN.R-project.org/package=microbenchmark},
  }

@article{hyndman2006another,
  title={Another look at measures of forecast accuracy},
  author={Hyndman, Rob J and Koehler, Anne B},
  journal={International Journal of Forecasting},
  volume={22},
  number={4},
  pages={679--688},
  year={2006},
  publisher={Elsevier}
}

@article{makridakis2018m4,
  title={The M4 Competition: Results, findings, conclusion and way forward},
  author={Makridakis, Spyros and Spiliotis, Evangelos and Assimakopoulos, Vassilios},
  journal={International Journal of Forecasting},
  volume={34},
  number={4},
  pages={802--808},
  year={2018},
  publisher={Elsevier}
}

@article{spiliotis2019forecasting,
  title={Are forecasting competitions data representative of the reality?},
  author={Spiliotis, Evangelos and Kouloumos, Andreas and Assimakopoulos, Vassilios and Makridakis, Spyros},
  journal={International Journal of Forecasting},
  year={2019},
  publisher={Elsevier}
}

@article{cleveland1990stl,
  title={STL: a seasonal-trend decomposition procedure based on loess},
  author={Cleveland, Robert B and Cleveland, William S and McRae, Jean E and Terpenning, Irma},
  journal={Journal of Official Statistics},
  volume={6},
  number={1},
  pages={3--73},
  year={1990}
}

@article{kang2017visualising,
  title={Visualising forecasting algorithm performance using time series instance spaces},
  author={Kang, Yanfei and Hyndman, Rob J and Smith-Miles, Kate},
  journal={International Journal of Forecasting},
  volume={33},
  number={2},
  pages={345--358},
  year={2017},
  publisher={Elsevier}
}

@article{ingel2020correlated,
  title={Correlated daily time series and forecasting in the M4 competition},
  author={Ingel, Anti and Shahroudi, Novin and K{\"a}ngsepp, Markus and T{\"a}ttar, Andre and Komisarenko, Viacheslav and Kull, Meelis},
  journal={International Journal of Forecasting},
  volume={36},
  number={1},
  pages={121--128},
  year={2020},
  publisher={Elsevier}
}

@article{Petropoulos2014,
  title = {`{Horses} for Courses' in demand forecasting},
  author={Petropoulos, F and Makridakis, S and Assimakopoulos, V and Nikolopoulos, K},
  journal={European Journal of Operational Research},
  volume = {237},
  year={2014},
  number={1},
  pages={152--163}
}

@article{li2010flexible,
  title={Flexible modeling of conditional distributions using smooth mixtures of asymmetric student t densities},
  author={Li, Feng and Villani, Mattias and Kohn, Robert},
  journal={Journal of Statistical Planning and Inference},
  volume={140},
  number={12},
  pages={3638--3654},
  year={2010},
  publisher={Elsevier}
}

@article{villani2009regression,
  title={Regression density estimation using smooth adaptive Gaussian mixtures},
  author={Villani, Mattias and Kohn, Robert and Giordani, Paolo},
  journal={Journal of Econometrics},
  volume={153},
  number={2},
  pages={155--173},
  year={2009},
  publisher={Elsevier}
}

@article{petropoulos2020simple,
  title={A simple combination of univariate models},
  author={Petropoulos, Fotios and Svetunkov, Ivan},
  journal={International Journal of Forecasting},
  volume={36},
  number={1},
  pages={110--115},
  year={2020},
  publisher={Elsevier}
}

@article{lichtendahl2013better,
  title={Is it better to average probabilities or quantiles?},
  author={Lichtendahl Jr, Kenneth C and Grushka-Cockayne, Yael and Winkler, Robert L},
  journal={Management Science},
  volume={59},
  number={7},
  pages={1594--1611},
  year={2013},
  publisher={INFORMS}
}

@Article{Li2020,
  author  = {Li, Xixi and Kang, Yanfei and Li, Feng},
  journal = {Expert System with Applications},
  title   = {Forecasting with time series imaging},
  year    = {2020},
  pages   = {113680},
  volume  = {160},
}

@article{petropoulos2018exploring,
  title={Exploring the sources of uncertainty: Why does bagging for time series forecasting work?},
  author={Petropoulos, Fotios and Hyndman, Rob J and Bergmeir, Christoph},
  journal={European Journal of Operational Research},
  volume={268},
  number={2},
  pages={545--554},
  year={2018}
}

@article{kourentzes2019another,
  title={Another look at forecast selection and combination: Evidence from forecast pooling},
  author={Kourentzes, Nikolaos and Barrow, Devon and Petropoulos, Fotios},
  journal={International Journal of Production Economics},
  volume={209},
  pages={226--235},
  year={2019}
}

@article{wang2021uncertainty,
  title={The uncertainty estimation of feature-based forecast combinations},
  author={Wang, Xiaoqian and Kang, Yanfei and Petropoulos, Fotios and Li, Feng},
  journal={Journal of the Operational Research Society},
  year={2021},
  publisher={Palgrave Macmillan}
}

@Article{Hyndman2008,
  author    = {Hyndman, Rob J and Khandakar, Yeasmin},
  journal   = {Journal of Statistical Software},
  title     = {Automatic time series forecasting: the forecast package for {R}},
  year      = {2008},
  number    = {3},
  pages     = {1--22},
  volume    = {26},
  publisher = {American Statistical Association},
}

@article{tashman2000out,
  title={Out-of-sample tests of forecasting accuracy: an analysis and review},
  author={Tashman, Leonard J},
  journal={International journal of forecasting},
  volume={16},
  number={4},
  pages={437--450},
  year={2000},
  publisher={Elsevier}
}

@article{aiolfi2006persistence,
  title={Persistence in forecasting performance and conditional combination strategies},
  author={Aiolfi, Marco and Timmermann, Allan},
  journal={Journal of Econometrics},
  volume={135},
  number={1-2},
  pages={31--53},
  year={2006},
  publisher={Elsevier}
}

@article{matsypura2018optimal,
  title={Optimal selection of expert forecasts with integer programming},
  author={Matsypura, Dmytro and Thompson, Ryan and Vasnev, Andrey L},
  journal={Omega},
  volume={78},
  pages={165--175},
  year={2018},
  publisher={Elsevier}
}

@article{arinze1997combining,
  title={Combining and selecting forecasting models using rule based induction},
  author={Arinze, Bay and Kim, Seung-Lae and Anandarajan, Murugan},
  journal={Computers \& Operations Research},
  volume={24},
  number={5},
  pages={423--433},
  year={1997},
  publisher={Elsevier}
}

@article{venkatachalam1999intelligent,
  title={An intelligent model selection and forecasting system},
  author={Venkatachalam, AR and Sohl, Jeffrey E},
  journal={Journal of Forecasting},
  volume={18},
  number={3},
  pages={167--180},
  year={1999},
  publisher={Wiley Online Library}
}

@article{prudencio2004meta,
  title={Meta-learning approaches to selecting time series models},
  author={Prud{\^e}ncio, Ricardo BC and Ludermir, Teresa B},
  journal={Neurocomputing},
  volume={61},
  pages={121--137},
  year={2004},
  publisher={Elsevier}
}

@article{wang2006characteristic,
  title={Characteristic-based clustering for time series data},
  author={Wang, Xiaozhe and Smith, Kate and Hyndman, Rob},
  journal={Data mining and knowledge Discovery},
  volume={13},
  number={3},
  pages={335--364},
  year={2006},
  publisher={Springer}
}

@inproceedings{kuck2016meta,
  title={Meta-learning with neural networks and landmarking for forecasting model selection},
  author={K{\"u}ck, Mirko and Crone, Sven F and Freitag, Michael},
  booktitle={2016 International Joint Conference on Neural Networks (IJCNN)},
  pages={1499--1506}
}

@article{adya2000application,
  title={An application of rule-based forecasting to a situation lacking domain knowledge},
  author={Adya, Monica and Armstrong, J Scott and Collopy, Fred and Kennedy, Miles},
  journal={International Journal of Forecasting},
  volume={16},
  number={4},
  pages={477--484},
  year={2000},
  publisher={Elsevier}
}

@article{lemke2010meta,
  title={Meta-learning for time series forecasting and forecast combination},
  author={Lemke, Christiane and Gabrys, Bogdan},
  journal={Neurocomputing},
  volume={73},
  number={10-12},
  pages={2006--2016},
  year={2010},
  publisher={Elsevier}
}

@inproceedings{widodo2013model,
  title={Model selection using dimensionality reduction of time series characteristics},
  author={Widodo, Agus and Budi, Indra},
  booktitle={International Symposium on Forecasting, Seoul, South Korea},
  year={2013}
}

\newpage
\appendix
\section{Visualisations of the meta-learner coefficients}\label{appendix}

\begin{figure}
\centering
\includegraphics[width=0.8\textwidth]{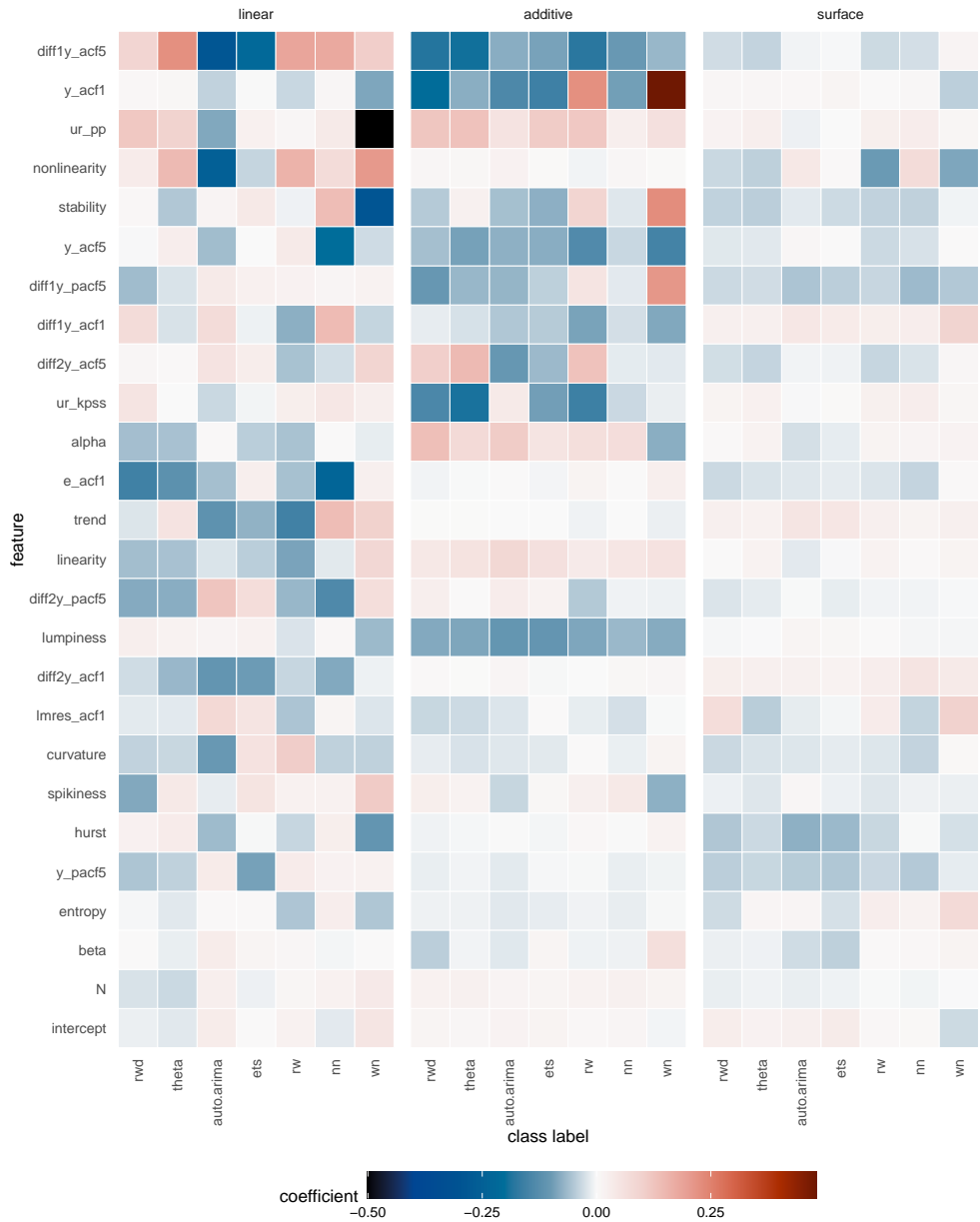}
  \caption{Visualisation of estimated model coefficients for yearly series. The Y-axis represents features, the X-axis represents class labels and the heat map cells represent estimated model coefficients. The class labels are ordered with their descending ranks.}
  \label{fig:ycoef}
\end{figure}

\begin{figure}
\centering
\includegraphics[width=0.9\textwidth]{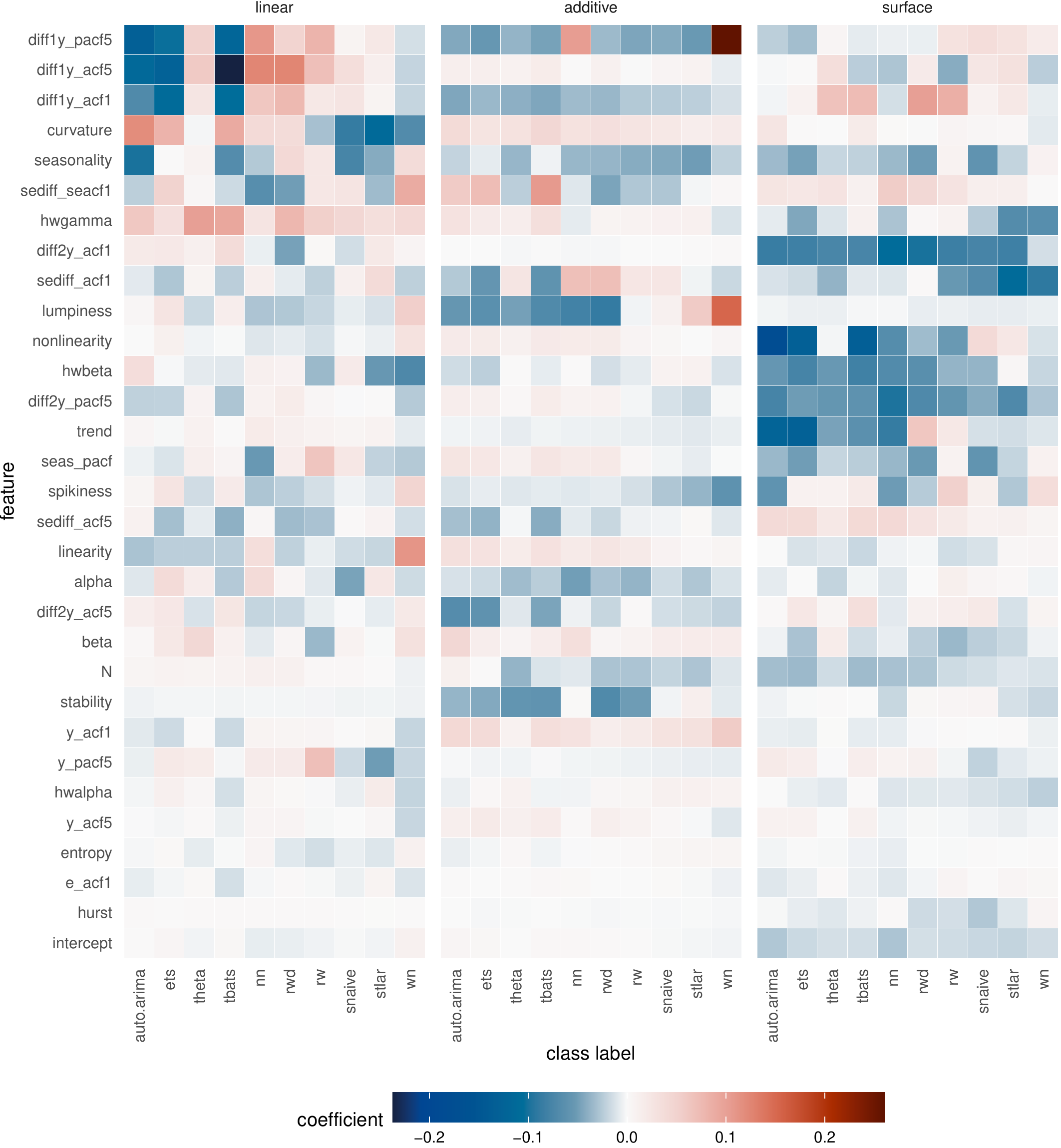}
  \caption{Visualisation of estimated model coefficients for monthly series. The Y-axis represents features, the X-axis represents class labels and the heat map cells represent estimated model coefficients. The class labels are ordered with their descending ranks.}
  \label{fig:mcoef}
\end{figure}

\begin{figure}
\centering
\includegraphics[width=\textwidth]{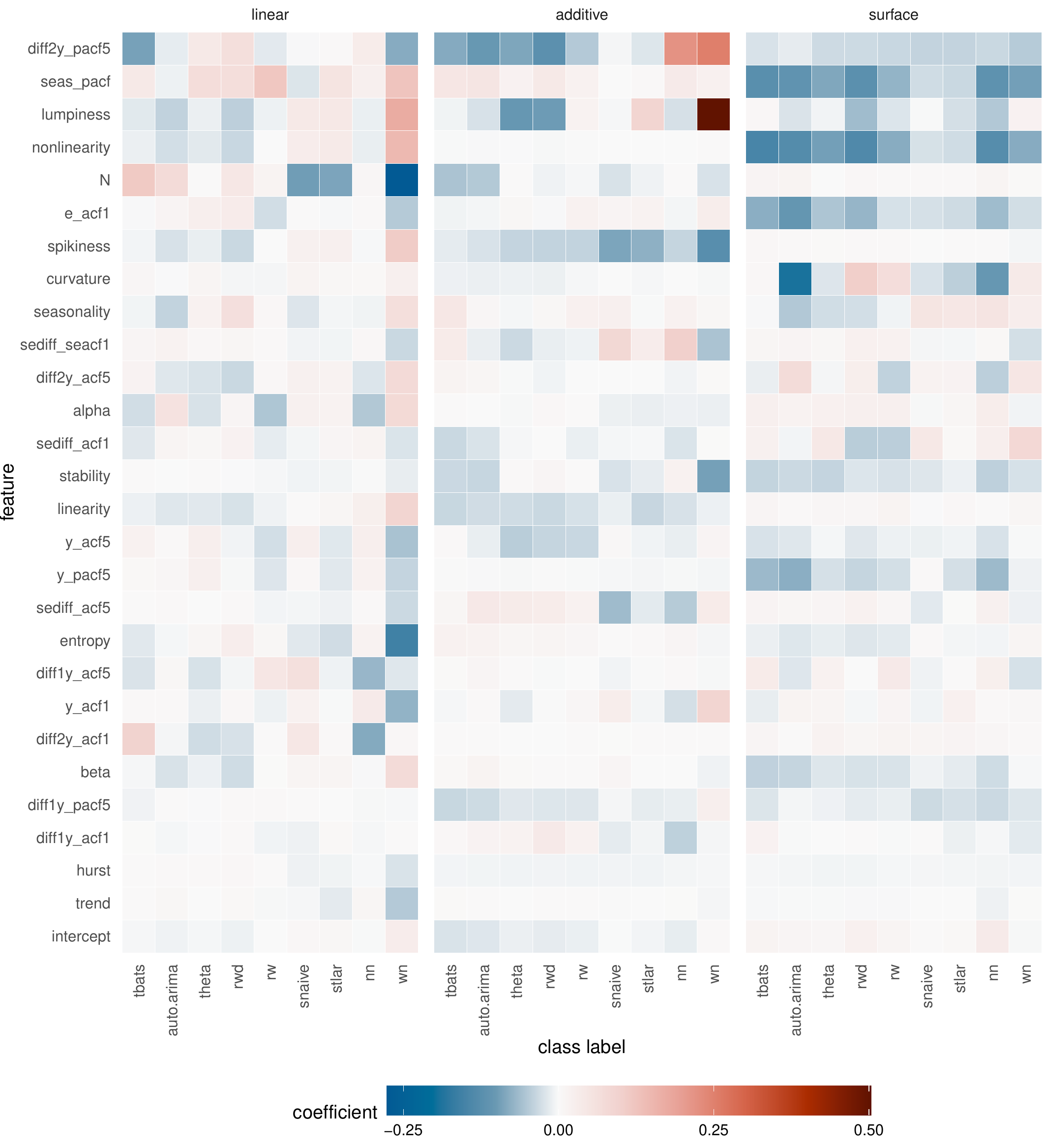}
  \caption{Visualisation of estimated model coefficients for weekly series. The Y-axis represents features, the X-axis represents class labels and the heat map cells represent estimated model coefficients. The class labels are ordered with their descending ranks.}
  \label{fig:wcoef}
\end{figure}

\begin{figure}
\centering
\includegraphics[width=\textwidth]{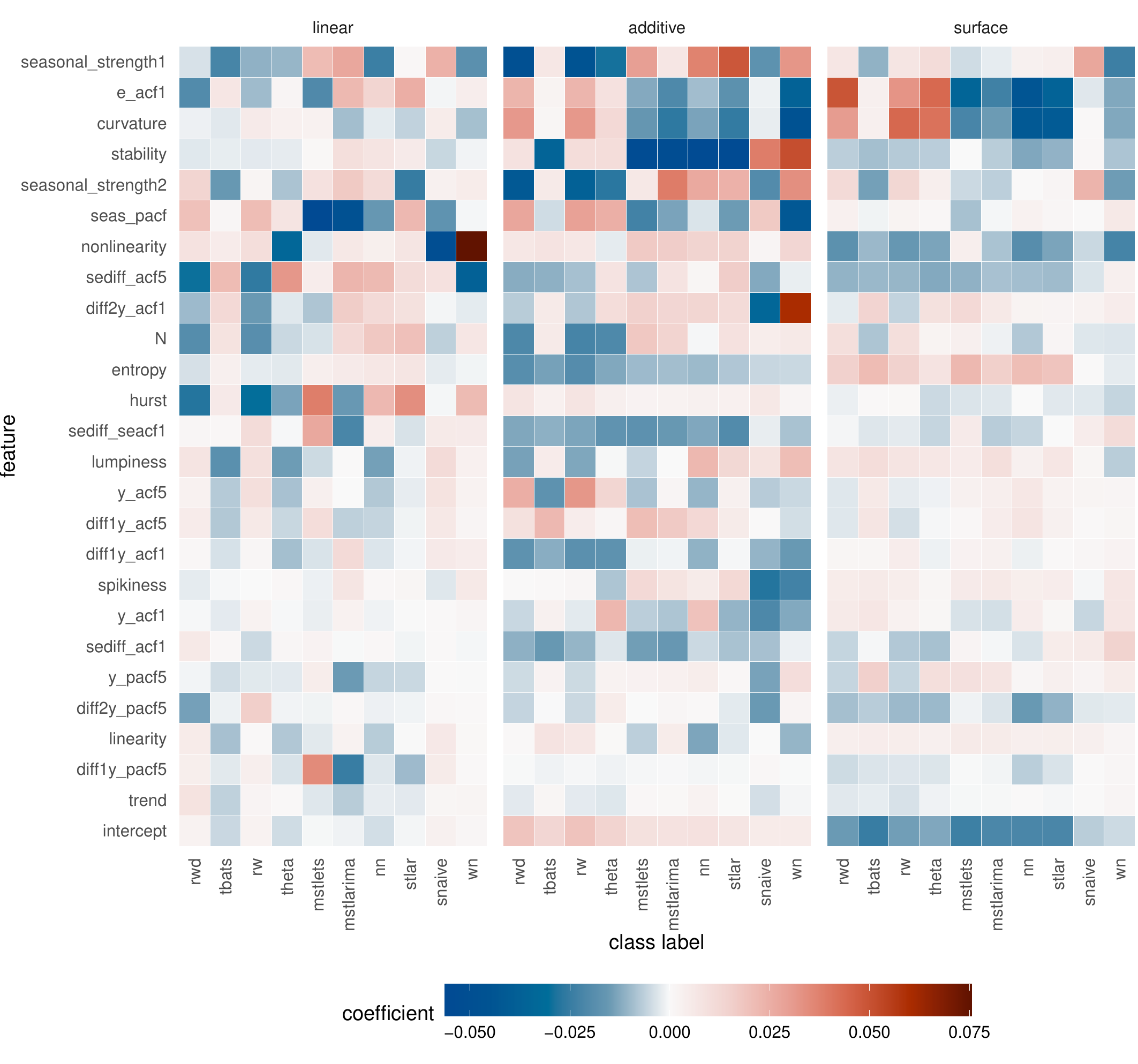}
  \caption{Visualisation of estimated model coefficients for daily series. The Y-axis represents features, the X-axis represents class labels and the heat map cells represent estimated model coefficients. The class labels are ordered with their descending ranks.}
  \label{fig:dcoef}
\end{figure}

\begin{figure}
\centering
\includegraphics[width=\textwidth]{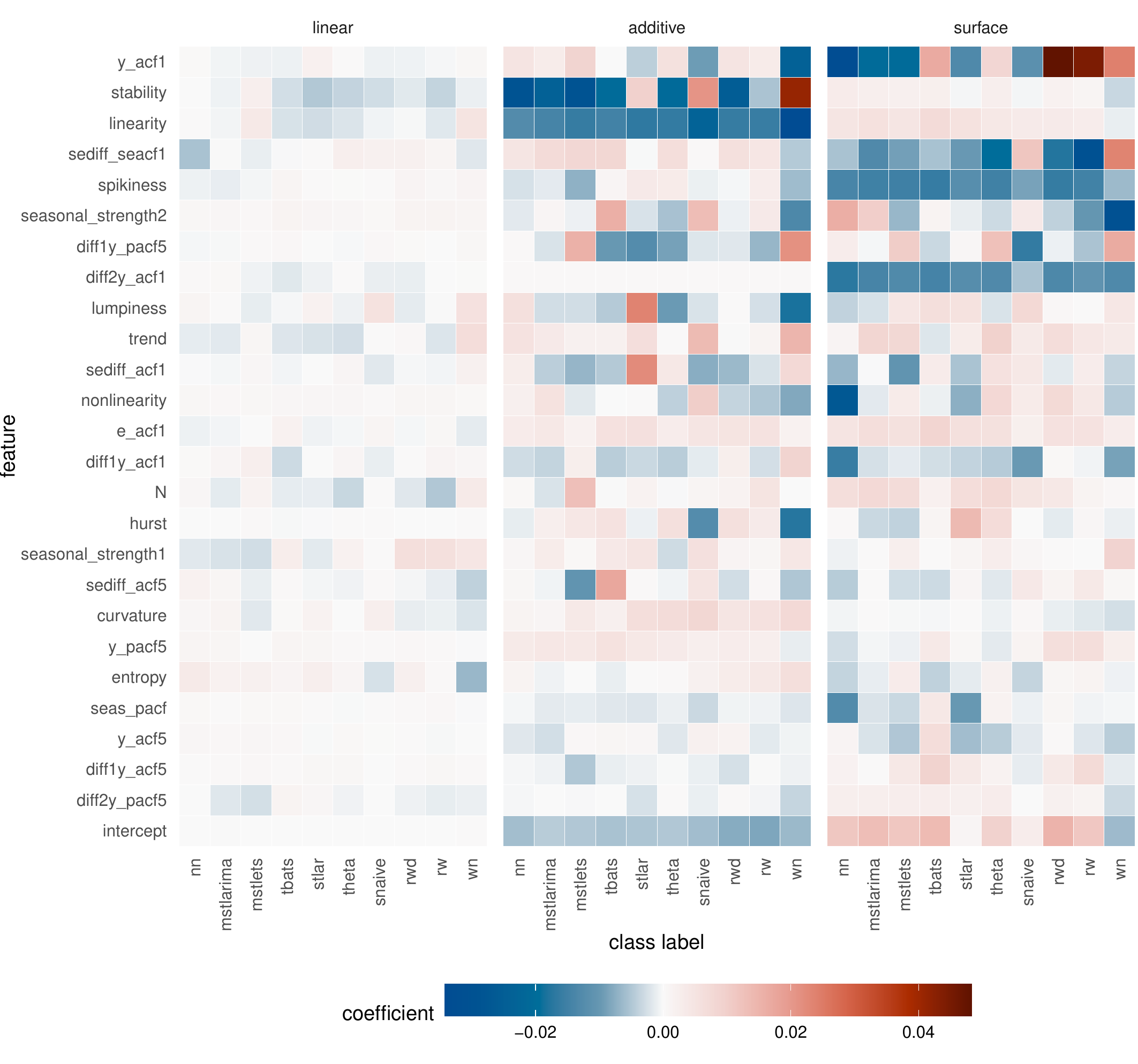}
  \caption{Visualisation of estimated model coefficients for hourly series. The Y-axis represents features, the X-axis represents class labels and the heat map cells represent estimated model coefficients. The class labels are ordered with their descending ranks.}
  \label{fig:hcoef}
\end{figure}

This section further visualises the model coefficients for the yearly (Figure
\ref{fig:ycoef}), monthly (Figure \ref{fig:mcoef}), weekly (Figure \ref{fig:wcoef}), daily
(Figure \ref{fig:dcoef}) and hourly (Figure \ref{fig:hcoef}) series. The heatmaps depict
that there are strong linear and additive nonlinear, but mild interactive (surface) nonlinear
relationships between features and the MASE vectors for the yearly data. For the monthly
and daily series, all three components are vital to the MASE vectors. Nonetheless, the
hourly series exhibit strong nonlinear characteristics. The contribution of features for
each the linear or nonlinear components is also observable from row-wise of the heatmaps.

\end{document}